# Shrinkage for Categorical Regressors[*]

Phillip Heiler[†]　　　Jana Mareckova[‡]

University of Konstanz　　University of Konstanz

October 26, 2018

This paper introduces a flexible regularization approach that reduces point estimation risk of group means stemming from e.g. categorical regressors, (quasi-)experimental data or panel data models. The loss function is penalized by adding weighted squared $\ell_2$-norm differences between group location parameters and informative first-stage estimates. Under quadratic loss, the penalized estimation problem has a simple interpretable closed-form solution that nests methods established in the literature on ridge regression, discretized support smoothing kernels and model averaging methods. We derive risk-optimal penalty parameters and propose a plug-in approach for estimation. The large sample properties are analyzed in an asymptotic local to zero framework by introducing a class of sequences for close and distant systems of locations that is sufficient for describing a large range of data generating processes. We provide the asymptotic distributions of the shrinkage estimators under different penalization schemes. The proposed plug-in estimator uniformly dominates the ordinary least squares in terms of asymptotic risk if the number of groups is larger than three. Monte Carlo simulations reveal robust improvements over standard methods in finite samples. Real data examples of estimating time trends in a panel and a difference-in-differences study illustrate potential applications.

[*]We would like to thank Lyudmila Grigoryeva, Chu-An Liu, Winfried Pohlmeier, Jeffrey S. Racine and the participants of the European Meeting of the Econometric Society 2016 for their time, comments, and discussions that helped to greatly improve the paper. The authors gratefully acknowledge the financial support from the German Research Foundation through research unit FOR 1882, Project 215899445 (Jana Mareckova) and Project 219805061 (Phillip Heiler) and by the Graduate School of Decision Sciences (University of Konstanz). All remaining errors are ours.

[†]Graduate School of Decision Sciences, Universitätsstraße 10, D-78462 Konstanz, Germany. Phone: +49-7531-88-5111, email: phillip.heiler@uni-konstanz.de.

[‡]Graduate School of Decision Sciences, Universitätsstraße 10, D-78462 Konstanz, Germany. Phone: +49-7531-88-2556, email: jana.mareckova@uni-konstanz.de.

# 1. Introduction

Estimation of conditional mean functions with categorical regressors can be very challenging. Even models with a moderate number of parameters lead to substantial estimation risk if the numbers of observations per group or cell that are determined by the explanatory variables are small. Regression models with multiple interactions, (quasi-)experimental designs or panel data models with time trends and fixed effects naturally fall into that category.

We propose a flexible penalization approach called pairwise cross-smoothing (PCS) to improve on the issue of point estimation risk. The method penalizes the loss function by adding sums of weighted squared $\ell_2$-norm differences between group location (reference) parameters and informative first-stage estimates (targets). It nests existing smoothing and averaging methods for orthogonal regressors, has favorable computational cost due to closed-form solutions for both estimator and penalty or smoothing parameters and can easily be extended to the case of mixed data.

Nonparametric methods in the fashion of Aitchison and Aitken (1976) are originally intended to deal with the small to empty cell problem in the context of multivariate discrete distributions (Hall, 1981; Simonoff, 1996) or mixed data distributions (Li and Racine, 2003; Hall et al., 2004). In the nonparametric regression framework Hall et al. (2007) and Ouyang et al. (2009) propose kernel methods with particular emphasis on cross-validated smoothing parameters and their behavior under the presence of irrelevant regressors. In a Bayesian sense, these methods shrink a multivariate mean towards a target value such as the global mean. The smoothing parameters depend only on a specific target covariate and are independent of the reference group. This is similar to (generalized) ridge regression (GRR) (Hoerl and Kennard, 1970). Smoothing a multivariate mean in the GRR context yields an optimization problem in which every location parameter $k$ is effectively shrunk towards a "leave the $k$-th group out average". Thus for any group, GRR effectively pushes the location parameter towards a joint target. In contrast to kernel regression, smoothing parameters depend only on the reference group and not the target. Pairwise cross-smoothing on the other hand allows nonhomogeneous smoothing for both reference and target groups. Therefore in the context of estimating group means, both kernel and (generalized) ridge regression can be seen as different restricted versions of PCS.

For probability distribution functions there is also a literature on empirical Bayes methods with data driven shrinkage parameters under appropriate priors for multinomial data, see e.g. Fienberg and Holland (1973), Titterington and Bowman (1985) or Simonoff (1995) for a comprehensive review with particular focus on sparse asymptotics.

The question of how to aggregate across distinctive groups can also be rephrased from a model or variable selection perspective, i.e. which groups deserve their own location parameter and which groups can be merged into one? In terms of a regression framework, one would like to know whether a more or less saturated model in terms of group dummy variables is appropriate. Classical model selection aims at selecting a single best model among a set of candidates by an appropriate criterion such as the Akaike Information Criterion (AIC, Akaike, 1973), Mallows $C_p$ (Mallows, 1973), the Schwarz-Bayes Criterion (BIC, Schwarz, 1978) or traditional multivariate testing procedures. There is no particular reason, why these discrete model selection approaches should always yield an optimal solution. In particular, if groups or parameters are different but close to each other, averaging parameter estimates across different models could serve as a superior model selection strategy. Hjort and Claeskens (2003) consider frequentist model averaging estimators and their distributional theory in a general maximum likelihood framework with a local to zero $n^{-1/2}$-asymptotic framework. See also Claeskens et al. (2008) for a comprehensive overview. Buckland et al. (1997) and Burnham and Anderson (2003) consider smooth variants of the AIC by applying exponential weighting structures. Hansen (2007) introduces a weighting procedure for least squares estimates based on Mallows Criterion. Liang et al. (2011) consider optimal weighting schemes in terms of the mean squared error for the linear model and general likelihood models.

Zhang et al. (2011) propose a focused information criterion and a model averaging estimator for generalized additive partially linear model with polynomial splines.

These smooth model averaging or shrinkage methods often have superior asymptotic risk properties over their non-shrunken counterparts. Hansen and Racine (2012) develop a jackknife model averaging estimator using cross-validation for conditional mean functions under potential misspecification of the submodels. They allow for heteroskedastic errors and non-nested models and show asymptotic optimality in the class of averaging estimators with weights in the unit simplex or a constrained subset thereof. Hansen (2014) derives conditions for asymptotic dominance of the averaging estimator in a nested least squares setup in a local to zero $n^{-1/2}$ framework, i.e. weak partial correlations of additional regressors beyond a correctly specified base model. Liu (2015) derives distributional theory for least squares averaging estimators in the linear framework under different data-dependent weighting schemes and generalized error term structures. He considers a local to zero $n^{-1/2}$-asymptotic framework for subsets of regressors and shows the nonstandard distributional behavior of the averaging estimators. Cheng et al. (2015) consider averaging between two general method of moments estimators under potential misspecification of the second, overidentified model. They show that the averaging estimator using estimated mean squared error optimal weights can dominate the asymptotic risk of the base estimator uniformly over all degrees of misspecification. Hansen (2016a) considers shrinkage of parametric models towards restricted parameter spaces under locally quadratic loss functions and provides conditions for risk dominance. If applied to cell means or selection of orthogonal dummies, many of these methods become special cases of PCS estimators, i.e. PCS estimators with a more restricted shrinkage subspace and thus behave qualitatively similar in terms of asymptotic distribution and estimation risk. They are also closely related to classical shrinkage estimators that shrink parametric estimates towards constant vectors or restricted subspaces (Stein, 1956; Oman, 1982).

In the literature on regularization methods, coefficient estimates are enhanced by adding $\ell_1$-norm penalties of pairwise differences which allows for partial and complete fusion of groups, see e.g. Tibshirani et al. (2005) for linear models and Tutz and Oelker (2017) for group-specific generalized linear models. The main differences to the other methods are nonsmooth aggregation, i.e. groups are set to be identical, and estimation that is done in a single, one-step procedure while e.g. model averaging directly and nonparametric smoothing implicitly use first-stage estimates such as submodels or averages. These regularization methods are more suited for sparse high-dimensional applications but suffer from similar criticism as pre-testing or superefficient estimators, i.e. in finite samples actual risk gains can be inferior to standard likelihood or least squares approaches and heavily depend on the size of the coefficients (Hansen, 2016b).

The direct or implicit aggregation that is introduced by all of these methods for regression models leads to the question of what an "optimal" aggregation rule is. Our framework allows for almost any linear aggregation based on a set of smoothing parameters. Using a simple projection as a first-stage estimate, we derive mean squared error optimal smoothing parameters and propose a plug-in approach for estimation. The additional flexibility provides a substantial decrease in the oracle risk bounds compared to more restrictive aggregation methods such as (generalized) ridge regression or kernel smoothing and thus also serves as a benchmark for future research.

We further contribute to the literature by analyzing the behavior of both estimated smoothing parameters as well as the PCS estimator in an asymptotic local to zero framework. We introduce a class of sequences for close and distant systems of locations that covers a wide range of data generating processes. We derive the asymptotic distribution of the PCS estimator under fixed, theoretically optimal and estimated smoothing parameters. In addition, we show that the feasible PCS dominates the OLS estimation risk uniformly over the class of sequences if the number of groups is larger than three.

Monte Carlo evidence suggests that the asymptotic uniform dominance property of the feasible PCS translates into superior finite sample performance over the OLS. More often than not, the method compares favorably to alternative shrinkage and model selection approaches such as (generalized) ridge regression, kernel smoothing and Mallows $C_p$.

The method is applied to the estimation of time trends in a short panel based on the field experiment in private day-care centers for children in Haifa by Gneezy and Rustichini (2000a) and to the difference-in-differences study about the effect of minimum wages on employment by Card and Krueger (1994) illustrating potential applications.

Section 2 introduces the model, the pairwise cross-smoothing estimator and its connection to established smoothing and regularization methods. Section 3 presents the MSE optimal smoothing parameters and the plug-in estimator. Section 4 introduces the local asymptotic framework and provides the distributional properties of the PCS estimator under fixed, optimal and plug-in weights. Section 5 discusses the asymptotic risk properties of the feasible PCS. Section 6 provides some Monte Carlo evidence on estimation risk in finite samples. Section 7 contains the applications. Section 8 concludes.

## 2. Pairwise Cross-Smoothing

### 2.1. The Model

In this section we introduce and discuss the model, the penalization strategy and the pairwise cross-smoothing estimator. Column vectors are denoted in boldface letters. Consider independent and identically distributed data $(Y_i, \mathbf{X}_i')$, $i = 1, \ldots, n$, where $Y_i$ is a real-valued random variable and $\mathbf{X}_i$ contains ordered and/or unordered discrete random variables[1]. These always uniquely determine $J$ orthogonal groups. For example, two binary discrete random variables determine four orthogonal groups. Let $\mathbf{D}_i$ indicate whether an observation $i$ belongs to a group $j \in \{1, \ldots, J\}$. In such a case, the $j$-th entry of the vector $\mathbf{D}_i$ contains one, $D_{ij} = 1$, and the remaining entries are equal to zero, $D_{ij'} = 0$ for all $j' \neq j$, i.e. $\mathbf{D}_i$'s have realizations in $\{\mathbf{e}_j, 1 \leq j \leq J\}$. We assume for the remainder that the groups are asymptotically non-empty, i.e. $P(D_{ij} = 1) \equiv p_j > 0$ for all $j$.

---

[1] For an extension to mixed data consider Online Appendix B.1.

Within this framework, a regression model for the conditional mean of $Y_i$ looks as follows:

$$Y_i = \mathbf{D}_i'\boldsymbol{\mu} + \varepsilon_i \tag{2.1}$$

with $\boldsymbol{\mu} = (\mu_1, \ldots, \mu_J)'$, $E[\varepsilon_i|\mathbf{D}_i] = 0$ and $V[\varepsilon_i|\mathbf{D}_i] = \sigma^2(\mathbf{D}_i)$ allowing for heteroskedasticity. Let $\hat{\boldsymbol{\mu}} = (\hat{\mu}_1, \ldots, \hat{\mu}_J)'$ be a consistent first-stage estimator for the group means. We propose to estimate the model for the conditional mean of $Y_i$ as a penalized least squares problem:

$$(\hat{\mu}_1^{PCS}, \ldots, \hat{\mu}_J^{PCS}) = \arg\min_{\mu_1, \ldots, \mu_J} \sum_{i=1}^{n}(Y_i - \mathbf{D}_i'\boldsymbol{\mu})^2 + Q(\boldsymbol{\Lambda}, \boldsymbol{\mu}, \hat{\boldsymbol{\mu}}) \tag{2.2}$$

$$Q(\boldsymbol{\Lambda}, \boldsymbol{\mu}, \hat{\boldsymbol{\mu}}) = \sum_{k=1}^{J}\sum_{j=1}^{J} \lambda_{kj}(\mu_k - \hat{\mu}_j)^2,$$

where $\boldsymbol{\Lambda} = (\lambda_{11}, \lambda_{12}, \ldots, \lambda_{1J}, \ldots, \lambda_{JJ})'$ are given smoothing or penalty parameters with $\lambda_{jj} = 0$ for all $j \in \{1, \ldots, J\}$. PCS stands for pairwise cross-smoothing since the penalties form hyperrectangles that geometrically overlap in $\mathbb{R}^{J-1}$. The idea behind the penalty is to improve the conditional group mean estimates by using information from other groups which is collected in the first-stage estimate. By allowing for reference and target dependent penalty parameters $\lambda_{kj}$, the penalty provides maximal flexibility for smoothing. The choice of smoothing

Regarding the choice of the smoothing parameters, the more informative group $j$ is for group $k$, the larger the smoothing parameter $\lambda_{kj}$ should be and vice versa. In the special case of $\lambda_{kj} = 0$ for all pairs $(k, j)$, none of the groups uses information from the other groups and the optimization is identical to the ordinary least squares problem. By choosing a large $\lambda_{kj}$, $\hat{\mu}_k^{PCS}$ is shrunk towards $\hat{\mu}_j$. Setting all $\lambda_{kj}$'s to large values pushes $\hat{\mu}_k^{PCS}$ towards the mean of all $\hat{\mu}_j$ where $j \neq k$. We discuss the issue of selecting optimal smoothing parameters in Section 3.

Let $n_k := \sum_{i=1}^{n} D_{ik}$ denote the number of observations within group $k$. Existence and uniqueness of the solution to (2.2) are guaranteed if $\sum_{l \neq k} \lambda_{kl} > -n_k$ for all $k \in \{1, \ldots, J\}^2$. Under this condition, the $k$-th group estimate is given by

$$\hat{\mu}_k^{PCS}(\mathbf{\Lambda}_k) = \frac{n_k \bar{Y}_k}{n_k + \sum_{l \neq k} \lambda_{kl}} + \sum_{j \neq k} \frac{\lambda_{kj} \hat{\mu}_j}{n_k + \sum_{l \neq k} \lambda_{kl}}, \qquad (2.3)$$

with $\mathbf{\Lambda}_k = (\lambda_{k1}, \ldots, \lambda_{kJ})'$ and $\bar{Y}_k$ denoting the sample mean of group $k$. One can see that the $k$-th group location estimator is a linear combination of its own cell mean and the first-stage group estimates.

A possible choice for $\hat{\boldsymbol{\mu}}$ is the linear (cell-based) projection of $Y_i$ on $\mathbf{D}_i$, i.e. $\hat{\boldsymbol{\mu}} = (\sum_{i=1}^{n} \mathbf{D}_i \mathbf{D}_i')^{-1} \sum_{i=1}^{n} \mathbf{D}_i Y_i$, the vector of cell means. The cell-based projection is also referred to as frequency approach in the literature since it weighs the outcomes only according to cell probabilities to form an estimate for a mean. The $k$-th mean estimator can then be written as:

$$\hat{\mu}_k^{PCS}(\mathbf{\Lambda}_k) = \frac{n_k \bar{Y}_k}{n_k + \sum_{l \neq k} \lambda_{kl}} + \sum_{j \neq k} \frac{\lambda_{kj} \bar{Y}_j}{n_k + \sum_{l \neq k} \lambda_{kl}}, \qquad (2.4)$$

which is a linear combination of cell means.

It is noteworthy that the smoothing parameters and therefore also the implicit weights $\lambda_{kj}/(n_k + \sum_{l \neq k} \lambda_{kl})$ are not all restricted to be larger or equal than zero. Just the overall smoothing for one reference category cannot be too negative. This fundamentally differentiates our approach from discretized support kernel approaches that are built as weighted averages using probability mass functions (Hall et al., 2004). They lead to weights which are restricted to be larger than zero. Shrinking simultaneously to different targets demands high flexibility from the smoothing parameters. Imposing strict positivity might not necessarily be optimal since smoothing away from distant groups can help to increase the smoothing to closer groups. The actual signs then depend on the absolute distances between group locations. For further

---

[2]For a complete proof see Appendix A.1.

discussion of the presence of negative smoothing parameters consider Section 3.

The penalty function can be considered a generalization of both generalized ridge regression (Hoerl and Kennard, 1970) and nonparametric kernel regression in the case of orthogonal binary regressors (Aitchison and Aitken, 1976; Ouyang et al., 2009). The generalized ridge estimator can be obtained by imposing equivalent shrinkage intensities $\lambda_{kj} = \lambda_k$ within all reference groups, i.e.

$$Q_{GRR}(\boldsymbol{\Lambda}, \boldsymbol{\mu}, \hat{\boldsymbol{\mu}}) = \sum_{k=1}^{J} \sum_{j \neq k} \lambda_k (\mu_k - \hat{\mu}_j)^2 \qquad (2.5)$$

$$\hat{\mu}_k^{GRR}(\lambda_k) = \frac{n_k \bar{Y}_k}{n_k + (J-1)\lambda_k} + \lambda_k \sum_{j \neq k} \frac{\hat{\mu}_j}{n_k + (J-1)\lambda_k}. \qquad (2.6)$$

Thus the GRR smooths every location parameter heterogeneously towards the corresponding shrinkage targets $\frac{1}{J-1}\sum_{j \neq k} \hat{\mu}_j$ that can be interpreted as "leave the $k$-th group out" averages.

The nonparametric smoothing kernel estimator can be obtained by imposing homogeneous shrinkage intensities $\lambda_{kj} = \lambda_j$ across all reference groups, i.e.

$$Q_{Kernel}(\boldsymbol{\Lambda}, \boldsymbol{\mu}, \hat{\boldsymbol{\mu}}) = \sum_{k=1}^{J} \sum_{j \neq k} \lambda_j (\mu_k - \hat{\mu}_j)^2 \qquad (2.7)$$

$$\hat{\mu}_k^{Kernel}(\boldsymbol{\Lambda}) = \frac{n_k \bar{Y}_k}{n_k + \sum_{l \neq k} \lambda_l} + \frac{\sum_{j \neq k} \lambda_j \hat{\mu}_j}{n_k + \sum_{l \neq k} \lambda_l}. \qquad (2.8)$$

Thus the estimator effectively smooths to a "weighted leave the $k$-th group out" average with homogeneous smoothing parameters for identical components across reference categories $k$.

A further restriction can be obtained via ordinary ridge regression with nonzero target by restricting the shrinkage intensities $\lambda_{kj} = \lambda$ to be equal for all reference groups and targets, i.e.

$$Q_{RR}(\boldsymbol{\Lambda}, \boldsymbol{\mu}, \hat{\boldsymbol{\mu}}) = \lambda \sum_{k=1}^{J} \sum_{j \neq k} (\mu_k - \hat{\mu}_j)^2. \qquad (2.9)$$

$$\hat{\mu}_k^{RR}(\lambda) = \frac{n_k \bar{Y}_k}{n_k + (J-1)\lambda} + \lambda \frac{\sum_{j \neq k} \hat{\mu}_j}{n_k + (J-1)\lambda}. \tag{2.10}$$

Thus ridge regression in this case smooths homogeneously towards the unweighted "leave the $k$-th group out" averages. It is reasonable to assume that allowing for more flexible shrinkage should be beneficial in terms of statistical risk if the smoothing parameters are chosen appropriately. This should be particularly pronounced if the groups are heterogeneous in terms of their size and variance.

Figure 2.1: Oracle Estimators - Relative Mean Squared Errors

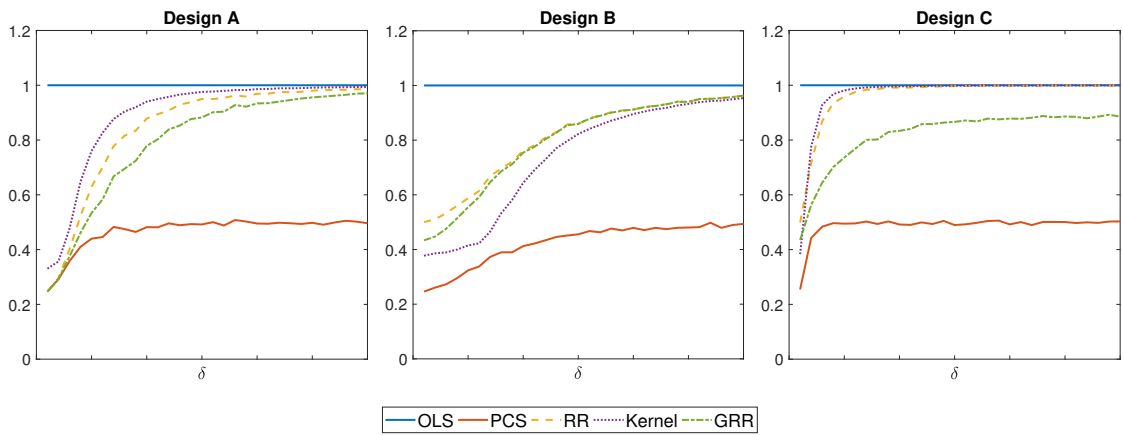

The figure depicts the simulated mean squared error relative to OLS for the estimators with risk-optimal (oracle) shrinkage parameters under normally distributed errors for $n = 400$, equal selection probabilities, and different parameter values $\delta$. The parameter vectors are $\boldsymbol{\mu}_A = \boldsymbol{\mu}_B = (0, 0, 0, \delta)'$, $\boldsymbol{\mu}_C = (0, 3\delta, -2\delta, \delta)'$. The group variances are $\boldsymbol{\sigma}_A^2 = (1, 1, 1, 1)'$ and $\boldsymbol{\sigma}_B^2 = \boldsymbol{\sigma}_C^2 = (1, 1, 1, 10)'$. Simulations are based on 5000 replications.

Figure 2.1 depicts the mean squared error of the different estimators relative to the OLS using mean squared error optimal smoothing parameters[3] in the case of four groups and different levels of heterogeneity regarding both means and variances. One can see that all of the approaches compare favorably to the OLS if the differences in group locations are not too big. The theoretically optimal PCS not only domi-nates all other approaches but also qualitatively behaves closer to a correctly chosen restricted estimator that has superior risk properties even when locations are not close to identical. In addition, for the very heterogeneous design C, the differences compared to the optimal kernel and optimal (G)RR are most pronounced. Thus the flexibility of the PCS penalty should in principle be able to generate substantial risk improvements compared to the alternative shrinkage methods.

---
[3] The PCS optimal smoothing parameters can be found in Section 3. For the alternative methods consider Appendix A.5.

## 3. Oracle Risk and Plug-In Estimation

In the following we provide a simple mean squared error criterion for evaluating the estimation risk, derive the optimal smoothing parameters and introduce a plug-in approach as a feasible counterpart. For the remainder we use a modified cell average (OLS) as a first-stage $\hat{\boldsymbol{\mu}}$ to assure existence, i.e.

$$\hat{\mu}_k = \frac{\sum_{i=1}^n D_{ik} Y_i}{\sum_{i=1}^n D_{ik} + \mathbb{1}(\sum_{i=1}^n D_{ik} = 0)} \tag{3.1}$$

for all $k$. Thus under given assumptions, a first order approximation for the first-stage is given by[4]

$$\hat{\mu}_k - \mu_k = \frac{1}{np_k} \sum_{i=1}^n D_{ik}(Y_i - \mu_k) + o_p(n^{-1/2}). \tag{3.2}$$

Note that rewriting (2.3) yields a weight-based representation of the PCS

$$\hat{\mu}_k^{PCS}(\boldsymbol{\Lambda}_k) \equiv \hat{\mu}_k^{PCS}(\boldsymbol{\omega}_k) = (1 - \sum_{j \neq k} \omega_{kj})\bar{Y}_k + \sum_{j \neq k} \omega_{kj}\hat{\mu}_j \tag{3.3}$$

with $\boldsymbol{\omega}_k = (\omega_{k1}, \ldots, \omega_{kJ})'$ and $\omega_{kj} = \lambda_{kj}/(n_k + \sum_{l \neq k} \lambda_{kl})$ being a one-to-one correspondence[5]. While the penalized regression representation is insightful for comparison with alternative methods, using the weighted version will simplify further

---

[4] See Appendix A.2.
[5] See Online Appendix B.2.

analysis. Using (3.2), the approximate parameter mean squared error is given by the following proposition:

**Proposition 3.1** *Let $E[Y_i|D_{ij} = 1] = \mu_j$, $V[\varepsilon_i|D_{ij} = 1] = \sigma_j^2$ be finite and $\hat{\boldsymbol{\mu}}$ be chosen according to (3.1). The MSE of the leading term of $\hat{\mu}_k(\boldsymbol{\omega}_k)$ is then given by*

$$MSE(\hat{\mu}_k^{PCS}(\boldsymbol{\omega}_k)) = \left(\sum_{j\neq k} \omega_{kj}(\mu_k - \mu_j)\right)^2 + (1 - \sum_{j\neq k} \omega_{kj})^2 \frac{\sigma_k^2}{np_k} + \sum_{j\neq k} \omega_{kj}^2 \frac{\sigma_j^2}{np_j}$$
$$= \boldsymbol{\omega}_k' \Delta_k \boldsymbol{\mu}\boldsymbol{\mu}' \Delta_k' \boldsymbol{\omega}_k + n^{-1} \boldsymbol{\omega}_k' diag(\boldsymbol{\gamma})^{-1} \boldsymbol{\omega}_k \qquad (3.4)$$

*with $\boldsymbol{\omega}_k = (\omega_{k1}, \ldots, \omega_{kJ})'$ s.t. $\boldsymbol{\omega}_k' \boldsymbol{\iota}_J = 1$, $\Delta_k$ being the k-th $J \times J$ dimensional partition of $\Delta = (I_J \otimes \boldsymbol{\iota}_J) - (\boldsymbol{\iota}_J \otimes I_J)$ and $\boldsymbol{\gamma} = (\gamma_1, \ldots, \gamma_J)'$ being vector of inverse OLS first stage variances with $\gamma_j = p_j/\sigma_j^2$.*

Note that due to the diagonal structure of the Gram matrix of the set of $\mathbf{D}_i$'s, the PCS estimator only depends on the smoothing parameters within its own reference category but is independent from the remaining smoothing parameters. Therefore, optimization of the parameter MSEs can be done group by group contrary to kernel smoothing and ridge regression. The following theorem establishes the MSE optimal smoothing parameters:

**Theorem 3.1** *The criterion in Proposition 3.1 is minimized at*

$$\omega_{kj}^* = \frac{\frac{p_j}{\sigma_j^2} + n\sum_{m\neq k}(\mu_k - \mu_m)(\mu_j - \mu_m)\frac{p_m p_j}{\sigma_m^2 \sigma_j^2}}{\sum_{l=1}^J \frac{p_l}{\sigma_l^2} + n\sum_{l=1}^J \sum_{m\neq k}(\mu_k - \mu_m)(\mu_l - \mu_m)\frac{p_l p_m}{\sigma_l^2 \sigma_m^2}}$$
$$= \frac{\gamma_j(1 + n\boldsymbol{\mu}'\Delta_k' diag(\boldsymbol{\gamma})\Delta_j \boldsymbol{\mu})}{\boldsymbol{\gamma}'\boldsymbol{\iota}_J + \frac{n}{2}\boldsymbol{\mu}'\Delta' M_1 \Delta \boldsymbol{\mu}} \qquad (3.5)$$

*with $M_1 = diag(\boldsymbol{\gamma}) \otimes diag(\boldsymbol{\gamma})$.*

The solution is always unique[6]. The minimizers corresponding to the $\lambda_{kj}$'s can be found in the Online Appendix. Note that if the first-stage estimates are all identical for a reference group, the corresponding optimal smoothing parameters become strictly positive. This is in line with Hoerl and Kennard (1970) who show for a generalized ridge regression that MSE optimal smoothing parameters have to be positive in the case of a common target. In the general case, negative smoothing parameters can be optimal due to different group specific targets.

---
[6]The optimal MSE smoothing parameters satisfy the existence and uniqueness condition for $\hat{\boldsymbol{\mu}}^{PCS}$. For more details consider Online Appendix B.3 and B.4.

The properties of PCS using oracle weights can be found in Section 4. One can construct the oracle weights for the restricted PCS, i.e. (G)RR and kernel regression in a similar fashion, however (weighted) MSE optimal ridge regression and kernel weights will in general not have a closed-form solution for larger $J$, see Appendix A.5.

While the oracle weights are theoretically appealing, they depend on unknown quantities through cell means, variances and probabilities and are generally infeasible. To construct a feasible counterpart we propose to replace the unknown quantities by consistent estimators. The plug-in weights are then given by

$$\hat{\omega}_{kj} = \frac{\frac{\hat{p}_j}{\hat{\sigma}_j^2} + n \sum_{m \neq k}(\hat{\mu}_k - \hat{\mu}_m)(\hat{\mu}_j - \hat{\mu}_m)\frac{\hat{p}_m \hat{p}_j}{\hat{\sigma}_m^2 \hat{\sigma}_j^2}}{\sum_{l=1}^{J} \frac{\hat{p}_l}{\hat{\sigma}_l^2} + n \sum_{l=1}^{J} \sum_{m \neq k}(\hat{\mu}_k - \hat{\mu}_m)(\hat{\mu}_l - \hat{\mu}_m)\frac{\hat{p}_l \hat{p}_m}{\hat{\sigma}_l^2 \hat{\sigma}_m^2}}$$
$$= \frac{\hat{\gamma}_j(1 + n\hat{\boldsymbol{\mu}}'\Delta_k' diag(\hat{\boldsymbol{\gamma}})\Delta_j \hat{\boldsymbol{\mu}})}{\hat{\boldsymbol{\gamma}}'\boldsymbol{\iota}_J + \frac{n}{2}\hat{\boldsymbol{\mu}}'\Delta'\hat{M}_1 \Delta \hat{\boldsymbol{\mu}}} \qquad (3.6)$$

with $\hat{p}_k = n_k/n$, $\hat{\sigma}_k^2 = \frac{1}{n_k-1}\sum_{i=1}^{n} D_{ik}(Y_i - \hat{\mu}_k)^2$ and equivalently for $\hat{\boldsymbol{\gamma}} = (\hat{\gamma}_1, \ldots, \hat{\gamma}_J)$, $\hat{\gamma}_k = \hat{p}_k/\hat{\sigma}_k^2$ and $\hat{M}_1 = diag(\hat{\boldsymbol{\gamma}}) \otimes diag(\hat{\boldsymbol{\gamma}})$ with $n_k \geq 2$ for all $k$. The feasible or plug-in PCS is then given by

$$\hat{\mu}_k^{PCS}(\hat{\boldsymbol{\omega}}) = \sum_{j=1}^{J} \hat{\omega}_{kj}\hat{\mu}_j \qquad (3.7)$$

The idea is that a first step is sufficiently informative for the optimal weights such that using a plug-in estimate will yield an estimated weighting scheme that improves on the actual performance of the resulting estimator. This approach is very close in spirit to other approaches based on MSE optimal averaging, focused information criteria and corresponding averaging estimators such as Hjort and Claeskens (2003), Liu (2015) and Cheng et al. (2015). In Section 5 we show that while oracle performance cannot be obtained for arbitrary data generating processes, the plug-in estimator still uniformly dominates the ordinary least squares in terms of (weighted) mean squared error.

Note that in contrast to the model averaging literature (Hansen and Racine, 2012; Liu, 2015) the weights are not restricted to lie in the unit simplex. Under fixed weights, the model averaging and the PCS estimator are linear in the outcome. For admissibility of linear estimators of the mean of a multivariate normal distribution, Cohen (1966) shows that symmetry and nonnegative eigenvalue bounds have to be met by the linear operator that maps outcomes to predictions, see also Li (1987) in a regression context. Hansen and Racine (2012) show that in the case of nested linear regression models, positivity of the model weights is a necessary condition for admissibility under mean squared error loss. However, if data dependent weights are used, the resulting estimator is no longer linear in the outcome. Furthermore, the less restrictive weighting scheme of the PCS can contradict the nesting requirement by Hansen and Racine (2012) despite the fact that the submodels are effectively linear. As a consequence, the overall shrinkage sum and not each shrinkage parameter is bounded from below and thus inadmissibility of the PCS does not follow. Interestingly the eigenvalue conditions of Cohen (1966) for admissibility still hold with probability one for both optimal and feasible PCS[7].

## 4. Large Sample Theory

### 4.1. Local Parameterization

In the following, we derive and discuss the large sample properties of the different weighting schemes and of the PCS estimator over a sufficient class of data generating processes relevant in the context of group means. In particular, we would like to distinguish between systems of locations in which the differences between group means are small (*close* systems) and large (*distant* systems) for a given sample size.

---

[7]See Online Appendix B.6.

For simplicity, instead of invoking standard assumptions to assure consistency and asymptotic normality of the first stage through moment conditions or similar, we start from a less rigorous point[8]. Let $\mathcal{F}$ be the set of distribution functions.

**Definition 1** *A sequence of data generating processes $\{F_n\}$ is close with local parameter $\boldsymbol{\delta}$ if*

$$\{F_n\} \in S(\boldsymbol{\delta}, V_0)$$
$$S(\boldsymbol{\delta}, V_0) = \{\{F_n\} : F_n \in \mathcal{F}, \sqrt{n}\Delta\boldsymbol{\mu} \to \Delta\boldsymbol{\delta} \in \mathbb{R}^{J^2}, \sqrt{n}(\hat{\boldsymbol{\mu}} - \boldsymbol{\mu}) \xrightarrow{d} \mathbf{Z}, \hat{V} \xrightarrow{p} V_0\}$$

**Definition 2** *A sequence of data generating processes $\{F_n\}$ is distant if*

$$\{F_n\} \in S(\infty, V_0)$$
$$S(\infty, V_0) = \{\{F_n\} : F_n \in \mathcal{F}, \sup_{k,j}\sqrt{n}|\mu_k - \mu_j| \to \infty, \sqrt{n}(\hat{\boldsymbol{\mu}} - \boldsymbol{\mu}) \xrightarrow{d} \mathbf{Z}, \hat{V} \xrightarrow{p} V_0\}$$

with $\boldsymbol{\delta} = (\delta_1, \delta_2, \ldots, \delta_J)'$, $\mathbf{Z} \sim \mathcal{N}(\mathbf{0}, V_0)$, $\hat{V} = diag(n\hat{\sigma}_j^2/n_j)$ and $V_0 = diag(\sigma_j^2/p_j)$. Close systems require that all scaled pairwise differences do not diverge, i.e. their differences depend on the local parameters $\delta_k - \delta_j$[9]. This nests the case in which all means are exactly identical and the local parameters are zero. For distant systems we require the scaled differences to go to infinity for at least one pair in the system. The union of these systems is sufficiently rich to describe a wide range of data generating processes.

To further motivate these classes of sequences and in particular the rate at which the differences converge to the local parameters, consider $J$ locations that are estimated via least squares. Assume that the asymptotic variances are known. Let $Z_n$ be a random variable that converges in distribution to a standard normal. A simple

---

[8] Note that standard regularity conditions usually imply asymptotic normality for estimated cell probabilities and variances as well. However, this is not required for any of the results in this and the following subsection.

[9] Note that since the system effectively depends only on differences in local parameters, constant shifts to $\boldsymbol{\delta}$ do not affect the analysis.

test for equality of two means $\mu_k$ and $\mu_j$ can be rewritten as follows:

$$T_n = \sqrt{n}\frac{\hat{\mu}_k - \hat{\mu}_j}{\sqrt{\frac{\sigma_k^2}{p_k} + \frac{\sigma_j^2}{p_j}}} = \sqrt{n}\frac{\mu_k - \mu_j}{\sqrt{\frac{\sigma_k^2}{p_k} + \frac{\sigma_j^2}{p_j}}} + Z_n \qquad (4.1)$$

Using the local parameterization it follows that

$$T_n(F_n) \xrightarrow{d} \mathcal{N}\left((\delta_k - \delta_j)\bigg/\sqrt{\sigma_k^2/p_k + \sigma_j^2/p_j}, 1\right) \text{ if } \{F_n\} \in S(\boldsymbol{\delta}, V_0), \qquad (4.2)$$

$$P(|T_n(F_n)| > c) \to 1 \text{ for all } c > 0 \qquad \text{if } \{F_n\} \in S(\infty, V_0). \qquad (4.3)$$

Therefore, depending on the local parameter difference $\delta_k - \delta_j$, one can obtain a small, moderate or even large mean of the test statistics' distribution. In the special case of $\delta_k - \delta_j$ being exactly equal to zero, the local parameterization does no longer affect the asymptotic distribution and classical inference can be conducted using the standard normal distribution. It is apparent that in any other case, choosing a model based on such a test might be misleading. If the local parameter is at a size that centers the limiting distribution e.g. around the critical value used for rejection of the null hypothesis, rejection would occur with probability one half. If this pretest is used for model selection, it is likely to suggest an underparameterized model that translates into higher parameter risk. The PCS estimator can be considered as a smooth variant of such a classical pretesting based estimator. Hence, we expect it to perform better exactly in these regions in which type-II errors are relatively large. Standard asymptotic analysis, however, will always favor the more parameterized model except if parameters are exactly equal. Thus, the approximations based on the local asymptotic framework should be closer to the actual finite sample behavior. The intuition can directly be translated to simultaneous tests of equality in locations for more than two groups[10].

---

[10]Online Appendix B.5 contains an example using a Wald test for equality of all means.

## 4.2. Distributional Theory

For investigation of the large sample properties of the different PCS, the behavior of the smoothing parameters along the sequences of DGPs is crucial. We consider PCS with weights $\omega_{kj}$ that correspond to fixed penalty parameters $\lambda_{kj}$ in (2.2), MSE optimal penalty $\omega_{kj}^*$ and plug-in parameters $\hat{\omega}_{kj}$. The following lemma demonstrates the behavior of the different weighting schemes in large samples.

**Lemma 4.1** *Let $\omega_{kj}^f$, $\omega_{kj}^*$ and $\hat{\omega}_{kj}$ denote the PCS weights in (3.7) corresponding to fixed, MSE optimal according to (3.5) and plugin solutions according to (3.6). Their limiting behavior along the local parameterization is then given by:*

$$\omega_{kj}^f = O_p(n^{-1}) \text{ if } k \neq j \text{ and } \omega_{kk}^f = 1 + O_p(n^{-1}) \qquad \text{if } \{F_n\} \in S(\boldsymbol{\delta}, V_0) \cup S(\infty, V_0)$$

$$\omega_{kj}^* \to \bar{w}_{kj} = \frac{\gamma_j(1 + \boldsymbol{\delta}'\Delta_k' diag(\boldsymbol{\gamma})\Delta_j\boldsymbol{\delta})}{\boldsymbol{\gamma}'\boldsymbol{\iota}_J + \frac{1}{2}\boldsymbol{\delta}'\Delta' M_1\Delta\boldsymbol{\delta}} \qquad \text{if } \{F_n\} \in S(\boldsymbol{\delta}, V_0)$$

$$\omega_{kj}^* \to \bar{w}_{kj} = 2\gamma_j \frac{\boldsymbol{\mu}'\Delta_k' diag(\boldsymbol{\gamma})\Delta_j\boldsymbol{\mu}}{\boldsymbol{\mu}'\Delta' M_1\Delta\boldsymbol{\mu}} \qquad \text{if } \{F_n\} \in S(\infty, V_0)$$

$$\hat{\omega}_{kj} \xrightarrow{d} w_{kj}^a = \frac{\gamma_j(1 + (\mathbf{Z}+\boldsymbol{\delta})'\Delta_k diag(\boldsymbol{\gamma})\Delta_j(\mathbf{Z}+\boldsymbol{\delta}))}{\boldsymbol{\gamma}'\boldsymbol{\iota}_J + \frac{1}{2}(\mathbf{Z}+\boldsymbol{\delta})'\Delta' M_1\Delta(\mathbf{Z}+\boldsymbol{\delta})} \qquad \text{if } \{F_n\} \in S(\boldsymbol{\delta}, V_0)$$

$$\hat{\omega}_{kj} \xrightarrow{p} \bar{w}_{kj} = 2\gamma_j \frac{\boldsymbol{\mu}'\Delta_k' diag(\boldsymbol{\gamma})\Delta_j\boldsymbol{\mu}}{\boldsymbol{\mu}'\Delta' M_1\Delta\boldsymbol{\mu}} \qquad \text{if } \{F_n\} \in S(\infty, V_0)$$

*with $\Delta_k$ being the $k$-th $J \times J$ dimensional partition of $\Delta = (I_J \otimes \boldsymbol{\iota}_J) - (\boldsymbol{\iota}_J \otimes I_J)$ and $\boldsymbol{\gamma} = (\gamma_1, \ldots, \gamma_J)'$ being vector of inverse OLS first stage variances with $\gamma_j = p_j/\sigma_j^2$.*

Note that the MSE optimal smoothing parameters do not in general vanish asymptotically, i.e. there is potential aggregation even in the limit. This, however, does not exclude the possibility to completely smooth out uninformative groups in large samples. It is qualitatively different from the smoothing kernel approach where uninformative, i.e. conditionally independent, regressors are always smoothed to a global average with smoothing parameters converging to their upper bound (Hall et al., 2004, 2007). The estimated smoothing parameters converge in distribution to a function of a normal random vector if groups are locally close while under a distant system, they converge in probability to the oracle parameters. Thus, adding a single distant parameter to a locally close system is sufficient to obtain convergence in probability. This is due to the fact that since the effective shrinkage target in the PCS are weighted leave the *k*-th group out averages. If the *k*-th group is the distant one, a weighted combination of the set of locally close locations will be distant enough to pin down the optimal weights in probability. If the *k*-th group is within the set of the locally close groups, the leave the *k*-th group out average will contain the distant group which is sufficient in large samples to distinguish the shrinkage target from the reference mean and thus lead to probabilistic convergence. Only in the case of all groups being locally close, the differences are not sufficient such that the limiting behavior is governed by a continuous function of a random normal vector. However, due to the rate of convergence of the estimated smoothing parameters in the case of distant systems, they will have an effect on the first order term determining the limiting distribution of the PCS under estimated smoothing parameters compared to the oracle distribution.

The following theorem establishes the distributional behavior of the different PCS variants.

**Theorem 4.1** *Let $\boldsymbol{\omega}_k^f$, $\boldsymbol{\omega}_k^*$ and $\hat{\boldsymbol{\omega}}_k$ denote the vector of fixed, MSE optimal according to (3.5) and plug-in weights according to (3.6) for the PCS. The asymptotic distributions of the PCS estimators are given by*

$$\sqrt{n}(\hat{\mu}_k^{PCS}(\boldsymbol{\omega}^f) - \mu_k - B_{1k}(\boldsymbol{\omega}^f)) \xrightarrow{d} Z_k \sim \mathcal{N}\left(0, \frac{\sigma_k^2}{p_k}\right) \quad \text{if } \{F_n\} \in S(\boldsymbol{\delta}, V_0) \cup S(\infty, V_0)$$

$$\sqrt{n}(\hat{\mu}_k^{PCS}(\boldsymbol{\omega}^*) - \mu_k - B_{2k}(\boldsymbol{\omega}^*)) \xrightarrow{d} \mathcal{N}\left(0, \sum_{j=1}^{J} \bar{\omega}_{kj}^2 \frac{\sigma_j^2}{p_j}\right) \quad \text{if } \{F_n\} \in S(\boldsymbol{\delta}, V_0) \cup S(\infty, V_0)$$

$$\sqrt{n}(\hat{\mu}_k^{PCS}(\hat{\boldsymbol{\omega}}) - \mu_k) \xrightarrow{d} \sum_{j=1}^{J} \omega_{kj}^a Z_j + \sum_{j=1}^{J} \omega_{kj}^a (\delta_j - \delta_k) \quad \text{if } \{F_n\} \in S(\boldsymbol{\delta}, V_0)$$

$$\sqrt{n}(\hat{\mu}_k^{PCS}(\hat{\boldsymbol{\omega}}) - \mu_k - B_{3k}(\bar{\boldsymbol{\omega}})) \xrightarrow{d} Z_k \sim \mathcal{N}\left(0, \frac{\sigma_k^2}{p_k}\right) \quad \text{if } \{F_n\} \in S(\infty, V_0)$$

with $B_{1k}(\boldsymbol{\omega}^f) = \sum_{j \neq k} \omega_{kj}^f (\mu_j - \mu_k)$, $B_{2k}(\boldsymbol{\omega}^*) = \sum_{j \neq k} w_{kj}^* (\mu_j - \mu_k)$, $B_{3k}(\bar{\boldsymbol{\omega}}) = \sum_{j \neq k} \bar{\omega}_{kj} (\mu_j - \mu_k)$.

Theorem 4.1 contains the asymptotic distributions of the different PCS estimators. There are some results that clearly parallel the literature on (generalized) ridge regression. In particular, a fixed penalty is asymptotically negligible for the distribution and thus the efficiency of the estimator, i.e. PCS with fixed weights converges in distribution to the corresponding OLS limit. The PCS under optimal weights differs in terms of its distribution from the OLS and thus the improvements in MSE in general do not disappear even for large samples. The behavior of the PCS under estimated smoothing parameters is particularly noteworthy. Note that the distribution to the normal is not uniform along all sequences of DGPs. In particular, the PCS with estimated smoothing parameters under locally close systems converges in distribution to a sum of normal random variables and local parameters multiplied by the limiting weights that are themselves functions of the same random normal variables, local parameters and other features of the DGP. Thus the limiting distribution in close systems is in general different from the normal. Assessing or estimating that limiting distribution has to be done with caution as the local parameters cannot be estimated consistently due to the $\sqrt{n}$ multiplier. This is similar to other shrinkage and model averaging methods that rely on smooth aggregation methods in the spirit of James-Stein shrinkage and frequentist model averaging (Hjort and Claeskens, 2003; Liu, 2015; Cheng et al., 2015; Hansen, 2016a).

## 5. Asymptotic Risk

Theorem 4.1 shows that when evaluating the risk of the feasible PCS, one has to take the additional variation of the weights under locally close systems into account. In the following we will focus on the risk under close systems as from there distant systems can be obtained as a special case by letting $||\Delta\boldsymbol{\delta}||_\infty \to \infty$. For derivation of the oracle risk, recall that due to the flexibility of the PCS, optimization can be done separately for each individual group. When evaluating the risk of the feasible PCS parameters however, the additional (co)variation introduced by the weighting parameters has to be taken into account since the latter are functions of the same random vector. The choice for the joint loss function will be the (weighted) parameter vector MSE

$$l(\tilde{\boldsymbol{\mu}}, \boldsymbol{\mu}) = (\tilde{\boldsymbol{\mu}} - \boldsymbol{\mu})'W(\tilde{\boldsymbol{\mu}} - \boldsymbol{\mu}) \tag{5.1}$$

with the canonical weighting matrix being the inverse of the asymptotic variance of the OLS or MLE parameter vector $W = diag(\boldsymbol{\gamma})$. Thus $W$ is proportional to the identity under homoskedasticity and equal group probabilities. The choice of $W$ also renders the evaluation of the risk invariant to rotations of the parameter vector, such that PCS risk properties are preserved even if outcomes are not generated on the same scale across groups.

To assure existence of a criterion that properly approximates the risk, a trimmed expected scaled loss criterion is used with vanishing trimming boundaries as in Hansen (2016a). Alternatively one could impose additional moment assumptions along the sequences to assure uniform integrability of the scaled loss function. In the limit, the risk is determined by a function of random normal vectors and thus easier to compute through the distributional limit. Let $\hat{\boldsymbol{\mu}}_n$ be a sequence of estimators along $\{F_n\} \in S(\boldsymbol{\delta}, V_0)$. The asymptotic risk along the sequences of DGPs is then given by

$$\rho(\hat{\boldsymbol{\mu}}_n, \boldsymbol{\mu}) = \lim_{\zeta \to \infty} \liminf_{n \to \infty} E_{F_n}[\min\{nl(\hat{\boldsymbol{\mu}}_n, \boldsymbol{\mu}), \zeta\}]. \tag{5.2}$$

Note that under normality, this would collaps to the exact finite sample risk. The risk of the OLS $\hat{\boldsymbol{\mu}}$ is then given by

$$\rho(\hat{\boldsymbol{\mu}}, \boldsymbol{\mu}) = tr(WV_0).$$

The following theorem provides the asymptotic risk of the PCS estimator under estimated weights for close systems of locations.

**Theorem 5.1** *Let $\{F_n\} \in S(\boldsymbol{\delta}, V_0)$. If $\hat{\boldsymbol{\omega}}$ is chosen according to (3.6), then*

$$\rho(\hat{\boldsymbol{\mu}}^{PCS}(\hat{\boldsymbol{\omega}}), \boldsymbol{\mu}) = E\left[\frac{(\mathbf{Z}+\boldsymbol{\delta})'\Delta'\{M_2 - tr(\Delta'M_3V_0)M_1 + 2M_3V_0\Delta'M_1\}\Delta(\mathbf{Z}+\boldsymbol{\delta})}{(tr(V_0^{-1}) + \frac{1}{2}(\mathbf{Z}+\boldsymbol{\delta})'\Delta'M_1\Delta(\mathbf{Z}+\boldsymbol{\delta}))^2}\right]$$
$$+ tr(WV_0) - 2tr(V_0^{-1})tr(\Delta'M_3V_0)E\left[\frac{1}{(tr(V_0^{-1}) + \frac{1}{2}(\mathbf{Z}+\boldsymbol{\delta})'\Delta'M_1\Delta(\mathbf{Z}+\boldsymbol{\delta}))^2}\right]$$

*with $V_0 = diag(\boldsymbol{\gamma})^{-1}$, $M_1 = diag(\boldsymbol{\gamma}) \otimes diag(\boldsymbol{\gamma})$, $M_2 = diag(\boldsymbol{\gamma}) \otimes \boldsymbol{\gamma}\boldsymbol{\gamma}'$ and $M_3 = diag(\boldsymbol{\gamma}) \otimes \boldsymbol{\gamma}$.*

Thus the asymptotic risk is given by the sum of the OLS risk, the expectation of the ratio of a quadratic form and a strictly positive random variable, and a strictly negative term. It depends on the limiting vector of the OLS, its asymptotic variance components and the unknown local parameter vector $\boldsymbol{\delta}$.

Theorem 5.1 allows us to establish sufficient conditions for a *strict uniform dominance* of the PCS estimator compared to the OLS. Uniform in the sense that for all bounded $\boldsymbol{\delta}$ vectors, the risk is strictly smaller than the risk of the OLS. It turns out that an easily interpretable sufficient condition is that the number of groups has to exceed three, i.e. we obtain the following corollary:

**Corollary 5.1** *Let $\{F_n\} \in S(\boldsymbol{\delta}, V_0)$ and $\hat{\boldsymbol{\omega}}$ be chosen according to (3.6). If $J \geq 4$, then*

$$\sup_{\boldsymbol{\delta} \in B} \rho(\hat{\boldsymbol{\mu}}^{PCS}(\hat{\boldsymbol{\omega}}), \boldsymbol{\mu}) - \rho(\hat{\boldsymbol{\mu}}, \boldsymbol{\mu}) < 0 \tag{5.3}$$

*for any bounded $B \subset \mathbb{R}^{J^2}$.*

Thus a sufficient (but by no means necessary) condition for uniform dominance is a simple condition on the dimensionality of the mean vector. This is similar to classical James-Stein estimation that as a necessary condition requires at least a three-dimensional multivariate mean vector when shrinking to a fixed target for global risk reduction over the maximum-likelihood estimator (Stein, 1956). Here, the somewhat more flexible shrinkage target requires one additional dimension, i.e. at least four groups to assure a strictly smaller risk for any close system of locations. Consistency follows directly as a corollary. In a similar spirit, the PCS shrinks towards a restricted subspace, i.e. an estimator that equalizes the group locations under a generalized error term structure. The corresponding subspace has exactly dimensionality $l = 1$ thus the minimal condition for superior risk (Oman, 1982) is that $J \geq 3 + l$ which equals the sufficient condition from Corollary 5.1.

## 6. Monte Carlo Study

The following simulations compare the small sample behavior of the PCS estimator to potential alternatives over a large range of data generating processes that vary with respect to mean parameters and error variances across groups. We investigate the weighted parameter vector MSE under close systems for different local parameter values $\delta$. The distant system behavior can be inferred for large values of $\delta$. The following estimators are considered:

1. Ordinary least squares/frequency method (OLS),
2. pairwise cross-smoothing with plug-in smoothing parameters (PCS),
3. ridge regression estimator with the plug-in smoothing parameter (RR),
4. generalized ridge regression estimator with plug-in smoothing parameters (GRR),
5. nonparametric smoothing kernel with plug-in smoothing parameters (Kernel),
6. a selection/pretesting estimator based on Mallows $C_p$[11] (Mallows).

We study a setup with a moderate number of groups, i.e. $J = 4$, that are selected with equal likelihood. The three designs A, B and C are set such that mean vectors converge to the origin but vary in the degree of deviations from the origin in finite samples. The mean vectors take following values $\boldsymbol{\mu}_A = (0, 0, 0, \delta/\sqrt{n})'$, $\boldsymbol{\mu}_B = (0, 0, -3\delta/\sqrt{n}, \delta/\sqrt{n})'$ and $\boldsymbol{\mu}_C = (0, 2\delta/\sqrt{n}, -3\delta/\sqrt{n}, \delta/\sqrt{n})'$ with $\delta$ varying over a positive grid starting at 0. For $\delta = 0$, we are in the case of identical means, i.e. a global average would be the most efficient estimator. Regarding the error term, we consider homoskedastic and heteroskedastic standardized log-normal distributions[12]. All results are based on 5000 simulations. For the plug-in weights we use the formulas from (3.6). In the homoskedastic designs, we change the variance estimator as if we knew in advance that the error is homoskedastic, i.e. all residuals are used to estimate a single variance parameter for all groups.

---

[11]We consider all possible submodels and choose the one with the lowest criterion value according to Mallows (1973). We also experimented with generalizations that are robust with respect to different error term structures. However, the classical $C_p$ seems to dominate all adaptations in our simulations and thus results are omitted.

Weighted mean squared errors relative to OLS for $n = 400$ are reported in Figure 6.1. Results for other sample sizes follow the same patterns and are available upon request. First, note that GRR, RR and $C_p$ estimator do not always outperform the OLS in the chosen setups. Only PCS and kernel estimators seem to uniformly dominate OLS. Depending on the value of $\delta$ one can get up to 30% improvement in the parameter vector MSE by using PCS over OLS. The largest benefits are obtained at lower values of $\delta$ as in these settings the shrinkage estimator can benefit from taking information from the other similar groups.

The kernel estimator follows virtually the same pattern as the PCS especially for larger $\delta$ parameters. For smaller $\delta$ values, however, kernel performs worse (up to 5 percentage points) due to its restricted flexibility when exploiting designs of almost equal means.

The RR generally performs well especially for small $\delta$ yielding MSE improvements up to 50% compared to OLS. In these designs, RR seems to benefit from the smaller amount of penalty parameters in comparison to PCS which has to deal with additional estimation noise. Nevertheless, RR loses its advantage in heterogeneous designs where the more flexible methods can further improve on the parameter vector MSE, i.e. under heteroskedasticity it even performs similar or slightly worse than OLS for moderate $\delta$ values in design A.

The GRR performs poorly in design A for large values of $\delta$ as it tends to shrink three groups to the rather distant leave the $k$-th group out targets in finite samples. Thus GRR estimation lacks robustness regarding the shrinkage target in asymmetric designs. It can perform substantially worse than ordinary least squares for a large range of $\delta$ values inflating the risk by more than 20%. Introducing a higher degree of symmetry around the origin in designs B and C helps GRR to shrink to correct targets and improves its performance. However, as these symmetries are usually unknown, we do not recommend the use of the plug-in GRR for shrinking categorical regressors in applied work.

---

[12] All results are robust with respect to the error distribution. Results for the normal distribution do not differ qualitatively and are available upon request.

The pretest estimator based on Mallows $C_p$ has its worst performance for moderate values of $\delta$ as in this range it is challenging for the model selection criterion to detect the optimal aggregation strategy. It often yields an underfitted model that introduces too much bias into the parameter estimates in line with the discussion in Section 4.1. This is particularly prominent in designs B and C with risk inflations of over 60% compared to OLS. With a higher degree of deviations from the origin (design C), the $C_p$ estimator performs worse than OLS over a larger range of $\delta$ parameters. However, for extreme values of $\delta$ parameters, it can perform better or close to OLS if there are risk gains from aggregating identical groups (design A and B). In practice, these mean differences are usually unknown and thus using the model selection criterion can be detrimental to the estimation. The presence of data generating processes for which model selection criteria yield inferior risk is a well-known phenomenon in the literature on model selection and post-selection risk, see e.g. Leeb and Pötscher (2008) among many others.

PCS on the other hand is virtually never worse than OLS, i.e. it shows uniformly dominant behavior in line with our results in Section 5. However, PCS is not always beating all the competitors over all the designs and $\delta$ values. For example, for small $\delta$ values GRR and RR are up to 20 percentage points superior profiting from an accurate shrinkage target in the design of almost equal means and less penalty parameters to estimate. For large values of $\delta$, the pretest estimator can perform better than PCS as the risk optimal model can be obtained through strict aggregation. PCS and kernel estimator show similar behavior and both show robustness in terms of different designs. From the two, PCS slightly outperforms the kernel estimator for smaller values of $\delta$.

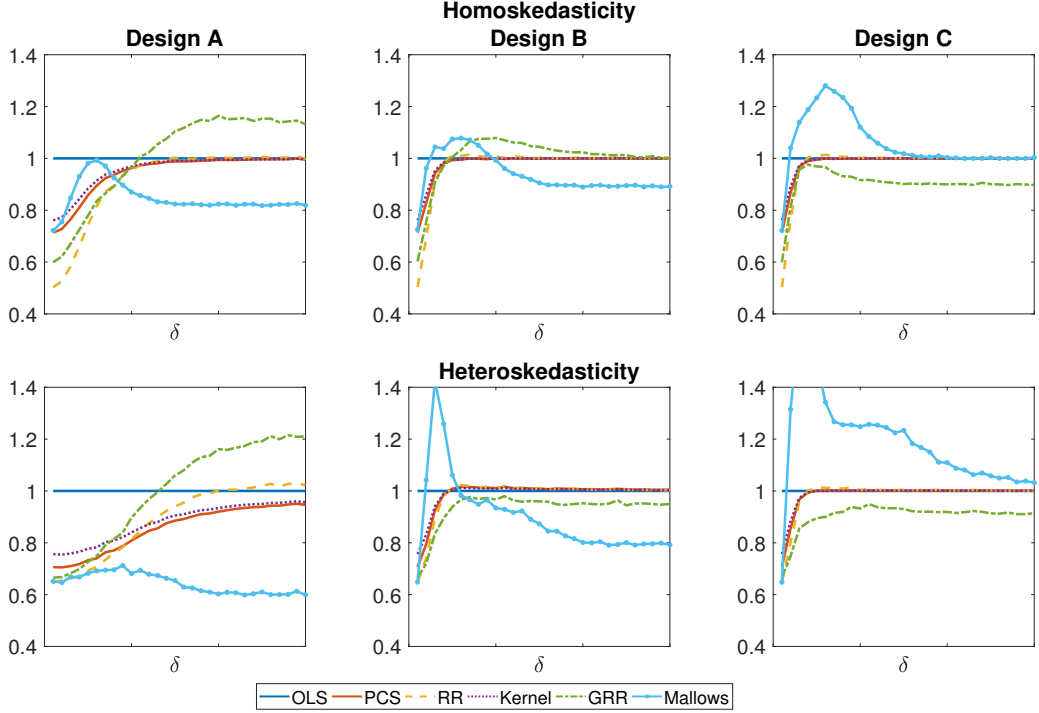

Figure 6.1: Relative Weighted Parameter Vector Mean Squared Errors

The figure depicts the simulated relative weighted parameter vector mean squared errors for the estimators with plug-in risk-optimal shrinkage parameters under log-normally distributed errors for $n = 400$, equal selection probabilities, different parameter values $\delta$, and different structures of the error variance with OLS as a benchmark. The parameter vectors are $\sqrt{n}\boldsymbol{\mu}_A = (0, 0, 0, \delta)'$, $\sqrt{n}\boldsymbol{\mu}_B = (0, 0, -3\delta, \delta)'$ and $\sqrt{n}\boldsymbol{\mu}_C = (0, 2\delta, -3\delta, \delta)'$. The group variances are $\boldsymbol{\sigma}^2_{hom} = (1, 1, 1, 1)'$ and $\boldsymbol{\sigma}^2_{het} = (1, 1, 1, 10)'$. Simulations are based on 5000 replications.

Therefore, PCS seems to be a robust refinement over OLS for a wide range of DGPs as alternatives are either dominated by PCS (OLS and kernel) or are very design sensitive (GRR and $C_p$). For the first category, PCS seems to be particularly superior for small values of $\delta$. For the latter category, there are always designs in which they perform (substantially) worse than OLS. Note that in general there is still room for further improvement since the large risk gains that can be obtained by the theoretically optimal PCS (see Section 2.1, Figure 2.1) cannot be reached by any method considered in the simulations designs.

# 7. Applications

## 7.1. Application I: A Fine is a Price

Gneezy and Rustichini (2000a) investigate the prediction of the deterrence hypothesis, i.e. that ceteris paribus introducing fines will decrease the likelihood of the associated action or behavior. They run a randomized control treatment study at ten day-care centers for young children in Haifa, Israel over a period of twenty weeks. It can be seen as a small panel data set with ten observations and twenty time periods. In period five, a fine is introduced for parents that come too late to pick up their children in six of these centers. They find that the fine increases the number of delayed parents and even after removal of the fine, the rate stayed at the same, higher level. The results have also been quoted in the literature on intrinsic and extrinsic motivation and crowding-out effects (Gneezy et al., 2011). Most of their major findings are summarized in a plot similar to the first subplot in Figure 7.1 which has been reused by e.g. Gneezy and Rustichini (2000b). In the variant used here, it depicts the share of late arrivals in both, treatment and control group over the duration of twenty weeks. Note that each point is an average over the subgroups of six and four data points in treatment and control group respectively which are basically predictions of a simple panel data model[13]. In statistical terms, it contains estimates for the expected share of late arrivals conditional on time period and treatment status. Our method is well-suited for this application since by construction, there are small orthogonal groups that are determined by time and treatment status. We stabilize the estimates of the conditional means by using the plug-in PCS within treatment groups and time periods closely related to Gneezy and Rustichini (2000a), Table 2. Hence we smooth the averages within weeks 1-4, 5-8, 9-16 and 17-20 for both groups using the original means as first stage[14].

Figure 7.1 depicts OLS (Gneezy and Rustichini, 2000a) and PCS estimates for the conditional mean over time and treatment status. The major findings of the original visualization are confirmed. In fact, our estimates reveal the pattern much clearer since the PCS suggests a more stable share of the control group and a less fluctuating mean of the treatment group before and after the time of treatment.

---

[13]Note that if only time trend dummy variables are used, a pooled OLS, fixed effect and random effect models coincide.

[14]This is based on the prior characterization of the periods by Gneezy and Rustichini (2000a). Of course other smoothing strategies could be employed as well.

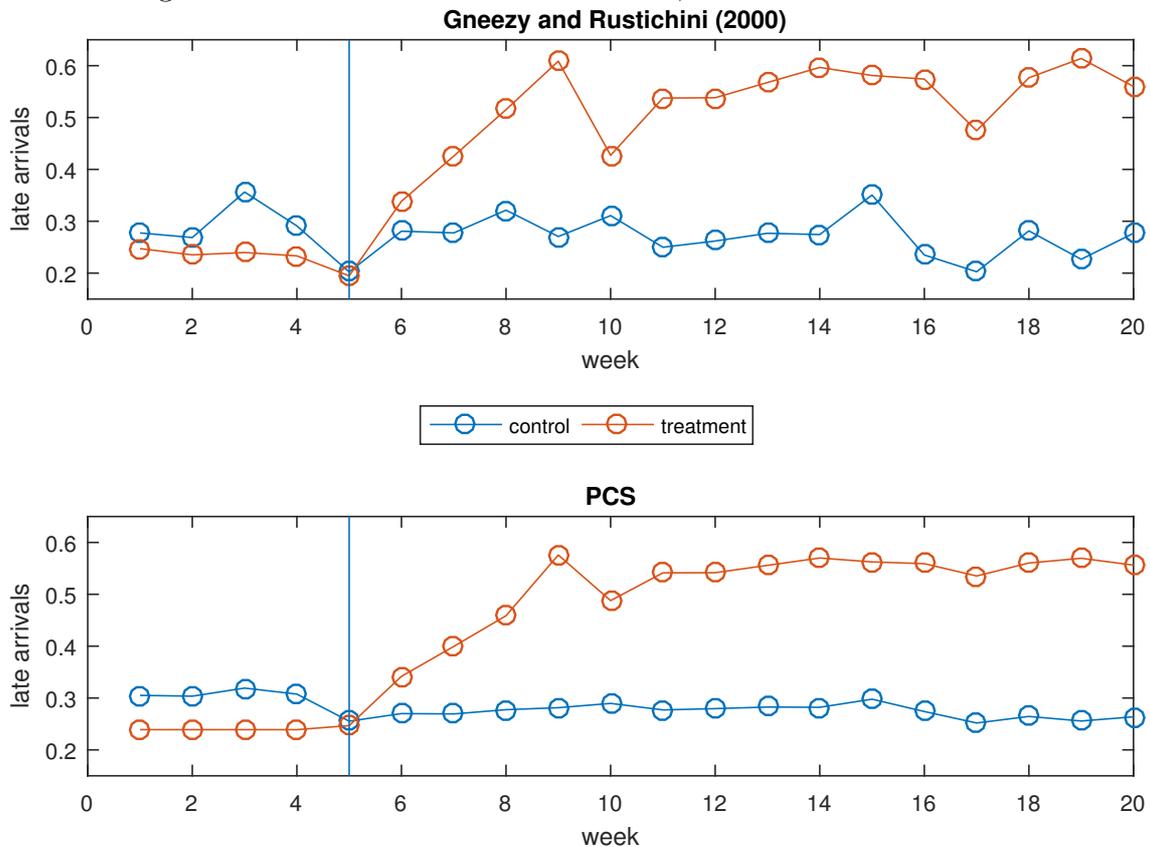

Figure 7.1: Mean Share of Late Arrivals, OLS and PCS estimates

### 7.2. Application II: Minimum Wage Study

The Card and Krueger (1994) paper is a case study evaluating the effects of minimum wage increase on the employment of low-wage workers. They collected data from fast food chains in New Jersey and Pennsylvania in a telephone survey before and after a minimum wage increase in New Jersey from 4.25\$ to 5.05\$ in 1992. The dependent variable full-time employment equivalent is measured as the number of full-time workers plus 0.5 times the part-time workers.

The setup is well-suited for our method since there are four orthogonal groups by construction that are determined by state and time. The PCS estimator is applied to the difference-in-differences model on the original Card and Krueger (1994) data and for each fast food chain separately to account for potentially different time trends and heterogeneous effects on employment across chains. As mentioned in Card and Krueger (1994), KFC differs in its size, opening hours and type of food from the other chains and therefore might be a source of heterogeneity. The chain by chain analysis further reduces the observations per cell and thus benefits the application of PCS over OLS. An alternative strategy would be to also smooth across chains to further increase the possible number of shrinkage targets.

The OLS (Card and Krueger, 1994) and PCS results for pooled data and for each chain separately are reported in Table 7.1[15]. We find a positive significant change in employment for the pooled data. As further analysis shows, this result is driven by the significant positive employment change in Burger King. The other chains have no significant change in employment. Comparing the results across chains, KFC shows a different pattern from the other stores, as KFC is the only chain with a point estimate that is in line with the theory of increasing labor demand in a less labor costly environment, however the estimate is statistically insignificant.

All the estimated effects of the minimum wage on the employment are closer to zero for the PCS in comparison to the OLS. In the case of pooled data, Burger King, KFC and Roys, the differences between OLS and PCS are not as large. However in case of Wendys, the chain with smallest number of observations in the data set, the difference is more pronounced, showing the stabilizing property of PCS in such scenarios.

---

[15]Table A.1 in Appendix A.9 includes means, variances and number of observations for all subgroups.

Table 7.1: Mean and Difference-in-Differences Estimates

| All Chains | OLS | | PCS | |
|---|---|---|---|---|
| | NJ (t) | PEN (c) | NJ (t) | PEN (c) |
| B | 20.44 | 23.33 | 20.53 | 22.87 |
| A | 21.03 | 21.17 | 21.01 | 21.12 |
| DiD | | 2.75* | | 2.22* |
| **Burger King** | OLS | | PCS | |
| | NJ (t) | PEN (c) | NJ (t) | PEN (c) |
| B | 22.16 | 29.42 | 22.25 | 29.06 |
| A | 23.63 | 26.22 | 23.63 | 26.06 |
| DiD | | 4.67** | | 4.38** |
| **KFC** | OLS | | PCS | |
| | NJ (t) | PEN (c) | NJ (t) | PEN (c) |
| B | 12.79 | 10.71 | 12.76 | 10.92 |
| A | 13.73 | 13.00 | 13.60 | 12.96 |
| DiD | | −1.35 | | −1.20 |
| **Roys** | OLS | | PCS | |
| | NJ (t) | PEN (c) | NJ (t) | PEN (c) |
| B | 23.14 | 19.74 | 22.99 | 19.80 |
| A | 21.73 | 15.81 | 21.68 | 16.12 |
| DiD | | 2.52 | | 2.37 |
| **Wendys** | OLS | | PCS | |
| | NJ (t) | PEN (c) | NJ (t) | PEN (c) |
| B | 22.08 | 24.12 | 22.43 | 23.46 |
| A | 23.40 | 22.10 | 23.10 | 22.44 |
| DiD | | 3.35 | | 1.69 |

The table contains the mean estimates of the full-time employment equivalents. B = before the minimum wage increase, A = after the minimum wage increase, NJ = New Jersey, PEN = Pennsylvania, t = treated, c = control. DiD formula: $(\hat{\mu}_{NJ,A} - \hat{\mu}_{NJ,B}) - (\hat{\mu}_{PEN,A} - \hat{\mu}_{PEN,B})$. p-values (* < 0.1, ** < 0.05) are computed using conventional confidence bounds.

## 8. Concluding Remarks

Pairwise cross-smoothing provides a unifying framework to analyze and compare smoothing methods for categorical data that nests different approaches from the literature on (generalized) ridge regression, nonparametric smoothing kernels and model averaging. It penalizes $\ell_2$ differences between estimation parameters and first-stage estimates. The estimator can be easily implemented with standard software packages using the closed-form solutions derived in this paper. It has favorable risk properties compared to the ordinary least squares and other commonly used approaches. For future research, relaxing the assumption of a fixed number of groups, i.e. allowing for $J$ to grow with the sample size with closeness restrictions that are related to sparsity in the sense of few different locations and alternative risk functions in the sense of (Hansen, 2016a) should be considered.

# A. Appendix

## A.1. FOC and SOC Conditions for (2.2)

Let $S_{\boldsymbol{\Lambda}}(\boldsymbol{\mu})$ denote the objective function in (2.2) where $\boldsymbol{\Lambda} = (\lambda_{11}, \lambda_{12}, \ldots, \lambda_{1J}, \lambda_{21}, \ldots, \lambda_{JJ})$ and $\lambda_{kk} = 0$ for all $k = \{1, \ldots, J\}$. Note that:

$$\frac{\partial S_{\boldsymbol{\Lambda}}(\boldsymbol{\mu})}{\partial \mu_k} = -2 \sum_{i=1}^n (Y_i - \mathbf{D}_i' \boldsymbol{\mu}) D_{ik} + 2 \sum_{j=1}^J \lambda_{kj}(\mu_k - \hat{\mu}_j) \tag{A.1}$$

$$\frac{\partial^2 S_{\boldsymbol{\Lambda}}(\boldsymbol{\mu})}{\partial \mu_k^2} = 2 n_k + 2 \sum_{j=1}^J \lambda_{kj} \tag{A.2}$$

$$\frac{\partial^2 S_{\boldsymbol{\Lambda}}(\boldsymbol{\mu})}{\partial \mu_k \partial \mu_l} = 0 \qquad\qquad l \neq k \tag{A.3}$$

Setting (A.1) equal to zero to solve for $\hat{\mu}_k^{PCS}$ and rearranging the terms yields

$$\sum_{i=1}^n Y_i D_{ik} + \sum_{j=1}^J \lambda_{kj} \hat{\mu}_j = \hat{\mu}_k^{PCS}(\sum_{j=1}^J \lambda_{kj} + n_k).$$

The estimate $\hat{\mu}_k^{PCS}$ exists if and only if $\sum_{j=1}^J \lambda_{kj} \neq -n_k$.

The matrix of second derivatives of $S_{\boldsymbol{\Lambda}}(\boldsymbol{\mu})$ is a diagonal matrix that leads to a strictly convex penalty if and only if

$$\sum_{j=1}^J \lambda_{kj} > -n_k \text{ for all } k \in \{1, \ldots, J\}.$$

An estimator defined as the solution to (2.2) exists and is a unique global minimizer if and only if $\sum_{j=1}^J \lambda_{kj} > -n_k$ for all $k \in \{1, \ldots, J\}$.

## A.2. First Order Approximation of the Modified Least Squares

$$\hat{\mu}_k = \frac{\sum_{i=1}^n D_{ik} Y_i}{\sum_{i=1}^n D_{ik} + \mathbb{1}(\sum_{i=1}^n D_{ik} = 0)}$$

By definition

$$\hat{\mu}_k - \mu_k = \frac{\sum_{i=1}^n D_{ik}(Y_i - \mu_k)}{\sum_{i=1}^n D_{ik} + \mathbb{1}(\sum_{i=1}^n D_{ik} = 0)} - \mu_k \frac{\mathbb{1}(\sum_{i=1}^n D_{ik} = 0)}{\sum_{i=1}^n D_{ik} + \mathbb{1}(\sum_{i=1}^n D_{ik} = 0)}.$$

Using iid and $p_k = E[D_{ik}] > 0$, the WULLN and CLT imply

$$n^{-1} \sum_{i=1}^n D_{ik} = p_k + o_p(1)$$

$$n^{-1/2} \sum_{i=1}^n D_{ik}(Y_i - \mu_k) \xrightarrow{d} \mathcal{N}(0, \sigma_k^2 p_k).$$

Since $p_k > 0$ and means and the indicator are bounded a.s we have that

$$n^{-1/2} \mu_k \mathbb{1}(\sum_{i=1}^n D_{ik} = 0) = o_p(1)$$

and

$$n^{-1/2} \frac{\sum_{i=1}^n D_{ik}(Y_i - \mu_k)}{n^{-1} \sum_{i=1}^n D_{ik} + n^{-1}\mathbb{1}(\sum_{i=1}^n D_{ik} = 0)} = n^{-1/2} \frac{\sum_{i=1}^n D_{ik}(Y_i - \mu_k)}{p_k} + o_p(1)$$

which together imply that

$$\hat{\mu}_k - \mu_k = \frac{1}{np_k} \sum_{i=1}^n D_{ik}(Y_i - \mu_k) + o_p(n^{-1/2}).$$

## A.3. Proof of Proposition 3.1

The bias of the PCS using the first order approximation is then given by

$$E[\hat{\mu}_k^{PCS}(\boldsymbol{\omega}_k)] - \mu_k = \sum_{j=1}^J \omega_{kj}(E[\hat{\mu}_j] - \mu_k)$$

$$= \sum_{j=1}^J \omega_{kj}(\mu_j - \mu_k).$$

The variance of the approximated PCS is given by

$$V[\hat{\mu}_k^{PCS}(\boldsymbol{\omega}_k)] = \sum_{j=1}^{J} \omega_{kj}^2 V[\hat{\mu}_j]$$

as $E[\hat{\mu}_j] - \mu_j \approx 0$, observations being independent, and cell means being approximately uncorrelated, i.e.

$$E[(\hat{\mu}_k - \mu_k)(\hat{\mu}_l - \mu_l)] \approx \frac{1}{n^2 p_k p_l} \sum_{i=1}^{n} \sum_{j=1}^{n} E[D_{ik}(Y_i - \mu_k)] E[D_{jl}(Y_j - \mu_l)]$$

$$= 0$$

for all $k \neq l$. The variances are given by

$$V[\hat{\mu}_j] \approx E\left[\left(\frac{1}{np_j} \sum_{i=1}^{n} D_{ij}(Y_i - \mu_j)\right)^2\right]$$

$$= \frac{1}{np_j^2} E[D_{ij}(Y_i - \mathbf{D}_i'\boldsymbol{\mu})^2]$$

$$= \frac{1}{np_j^2} E[D_{ij}\varepsilon_i^2]$$

$$= \frac{1}{np_j} E[\varepsilon_i^2 | D_{ij} = 1]$$

$$= \frac{\sigma_j^2}{np_j}.$$

where second step is due to independent observations. The MSE in (3.4) then follows by the usual bias-variance decomposition.

## A.4. Proof of Theorem 3.1

The problem of the constrained minimization of (3.4) can be rewritten as the following Lagrangian:

$$\min_{\boldsymbol{\omega}_k, \alpha_k} \mathcal{L}(\boldsymbol{\omega}_k, \alpha_k) = \min_{\boldsymbol{\omega}_k, \alpha_k} \boldsymbol{\omega}'_k H_k \boldsymbol{\omega}_k + 2\alpha_k(1 - \boldsymbol{\iota}'_J \boldsymbol{\omega}_k),$$

$$\text{with } \boldsymbol{\omega}_k = (\omega_{k1}, \omega_{k2}, \ldots, \omega_{kJ})',$$

$$H_k = \Delta_k \boldsymbol{\mu} \boldsymbol{\mu}' \Delta'_k + diag(\boldsymbol{\gamma})^{-1}/n,$$

$$\alpha_k = \text{ Lagrange multiplier.}$$

The FOCs are given by

$$\frac{\partial \mathcal{L}(\boldsymbol{\omega}_k, \alpha_k)}{\partial \boldsymbol{\omega}_k} = 2H_k \boldsymbol{\omega}_k - 2\alpha_k \boldsymbol{\iota}_J = 0$$

$$\frac{\partial \mathcal{L}(\boldsymbol{\omega}_k, \alpha_k)}{\partial \alpha_k} = 2(1 - \boldsymbol{\iota}'_J \boldsymbol{\omega}_k) = 0$$

The solution of setting the FOC to zero gives optimal values:

$$\alpha_k^* = [\boldsymbol{\iota}'_J (H'_k H_k)^{-1} H'_k \boldsymbol{\iota}_J]^{-1}$$

$$\boldsymbol{\omega}_k^* = [\boldsymbol{\iota}'_J (H'_k H_k)^{-1} H'_k \boldsymbol{\iota}_J]^{-1} (H'_k H_k)^{-1} H'_k \boldsymbol{\iota}_J$$

The expression for $\omega_{kj}^*$ can be inferred from the $j$-th entry of $\boldsymbol{\omega}_k^*$. For uniqueness conditions consider Online Appendix B.3.

## A.5. WMSE Optimal and Plugin Smoothing Parameters for Kernel and (generalized) Ridge Regression

For all methods we choose the weighted MSE criterion with $W = V_0^{-1}$ being the inverse of the MLE/least squares variance-covariance matrix. The first-stage estimate is chosen to be the (modified) ordinary least squares. Note that for a given PCS estimator (or restricted version thereof) $\hat{\boldsymbol{\mu}}(\boldsymbol{\Lambda})$ with smoothing parameter vector

$\boldsymbol{\Lambda}$, the criterion can be written as:

$$E[(\hat{\boldsymbol{\mu}}(\boldsymbol{\Lambda}) - \boldsymbol{\mu})'W(\hat{\boldsymbol{\mu}}(\boldsymbol{\Lambda}) - \boldsymbol{\mu})] = \sum_{k=1}^{J} \gamma_k E[(\hat{\mu}_k(\boldsymbol{\Lambda}) - \mu_k)^2].$$

### A.5.1. Kernel Smoothing

The implicit kernel constraints yield the estimator of the form

$$\hat{\mu}_k^{Kernel}(\boldsymbol{\Lambda}) = \frac{n_k \bar{Y}_k}{n_k + \sum_{l \neq k} \lambda_l} + \frac{\sum_{j \neq k} \lambda_j \hat{\mu}_j}{n_k + \sum_{l \neq k} \lambda_l}. \tag{A.4}$$

or equivalently by using the WLLN and $\bar{Y}_k = \hat{\mu}_k$

$$\hat{\mu}_k^{Kernel}(\boldsymbol{\Lambda}) - \mu_k = \frac{np_k(\hat{\mu}_k - \mu_k) + \sum_{j \neq k} \lambda_j(\hat{\mu}_j - \mu_k)}{np_k + \sum_{l \neq k} \lambda_l} + o_p(1) \tag{A.5}$$

and thus

$$(\hat{\mu}_k^{Kernel}(\boldsymbol{\Lambda}) - \mu_k)^2 \approx \frac{(np_k)^2(\hat{\mu}_k - \mu_k)^2 + 2np_k(\hat{\mu}_k - \mu_k)\sum_{j \neq k} \lambda_j(\hat{\mu}_j - \mu_k)}{(np_k + \sum_{l \neq k} \lambda_l)^2}$$
$$+ \frac{\sum_{j \neq k}\sum_{m \neq k,j} \lambda_j \lambda_m (\hat{\mu}_j - \mu_k)(\hat{\mu}_m - \mu_k) + \sum_{j \neq k} \lambda_j^2 (\hat{\mu}_j - \mu_k)^2}{(np_k + \sum_{l \neq k} \lambda_l)^2}.$$

$$\tag{A.6}$$

Since the groups means are uncorrelated, the expected risk for group $k$ is given by

$$E[(\hat{\mu}_k^{Kernel}(\boldsymbol{\Lambda}) - \mu_k)^2] \approx \frac{(np_k)^2 \frac{1}{\gamma_k n} + \sum_{j \neq k}\sum_{m \neq k,j} \lambda_j \lambda_m (\mu_j - \mu_k)(\mu_m - \mu_k)}{(np_k + \sum_{l \neq k} \lambda_l)^2}$$
$$+ \frac{\sum_{j \neq k} \lambda_j^2 [\frac{1}{\gamma_j n} + (\mu_j - \mu_k)^2]}{(np_k + \sum_{l \neq k} \lambda_l)^2} \tag{A.7}$$

which yields the overall weighted MSE

$$\sum_{k=1}^{J} \gamma_k E[(\hat{\mu}_k^{Kernel}(\boldsymbol{\Lambda}) - \mu_k)^2] \approx \sum_{k=1}^{J} \left[\frac{np_k}{np_k + \sum_{l \neq k} \lambda_l}\right]^2 \frac{1}{n} + \frac{\sum_{k=1}^{J} \sum_{j \neq k} \lambda_j^2 \frac{\gamma_k}{\gamma_j} \frac{1}{n}}{(np_k + \sum_{l \neq k} \lambda_l)^2}$$
$$+ \frac{\sum_{k=1}^{J} \sum_{j \neq k} \sum_{m \neq k} \gamma_k \lambda_j \lambda_m (\mu_j - \mu_k)(\mu_m - \mu_k)}{(np_k + \sum_{l \neq k} \lambda_l)^2}. \quad (A.8)$$

The plug-in estimator can be obtained by replacing the expression for $p_k$, $\gamma_k$ and $\mu_k$ for $k = 1 \ldots, J$ by the corresponding estimates as for the PCS and optimizing with respect to $\boldsymbol{\Lambda}$ using numerical optimization. Similar to kernel estimation for continuous data, there is no closed-form solution for the general case.

### A.5.2. Ridge Regression

For the ridge regression (RR), the restrictions imposed on the smoothing parameters are $\lambda_{kj} = \lambda$. This yields an estimator:

$$\hat{\mu}_k^{RR}(\lambda) = \frac{n_k}{n_k + (J-1)\lambda} \bar{Y}_k + \frac{(J-1)\lambda}{n_k + (J-1)\lambda} \sum_{j \neq k} \frac{\hat{\mu}_j}{J-1}.$$

Under the choice of $\hat{\mu}_j = \bar{Y}_j$, the weighted MSE of the leading term of a first-order approximation of the RR estimator takes form[16]:

$$MSE(\hat{\boldsymbol{\mu}}^{RR}(\lambda)) = E\left[(\hat{\boldsymbol{\mu}}^{RR}(\lambda) - \boldsymbol{\mu})'W(\hat{\boldsymbol{\mu}}^{RR}(\lambda) - \boldsymbol{\mu})\right] = E\left[\sum_{k=1}^{J} \frac{p_k}{\sigma_k^2}(\hat{\mu}_k^{RR}(\lambda) - \mu_k)^2\right]$$
$$\approx \sum_{k=1}^{J} \frac{p_k}{\sigma_k^2} \left(\left(\frac{(J-1)\lambda}{np_k + (J-1)\lambda}\right)^2 \left[\left(\sum_{j \neq k} \frac{\mu_j}{(J-1)} - \mu_k\right)^2 + \sum_{j \neq k} \frac{\sigma_j^2}{(J-1)^2 np_j}\right]\right.$$
$$\left. + \left(\frac{np_k}{np_k + (J-1)\lambda}\right)^2 \frac{\sigma_k^2}{np_k}\right).$$

---
[16]For more details about the first-order approximation please refer to Section 3.

The FOC for the weighted MSE optimal parameter $\lambda$ takes form:

$$\sum_{k=1}^{J} \frac{p_k}{\sigma_k^2} \left( \frac{\lambda(J-1)np_k}{(np_k + (J-1)\lambda)^3} \left[ \left( \sum_{j \neq k} \frac{\mu_j}{(J-1)} - \mu_k \right)^2 + \sum_{j \neq k} \frac{\sigma_j^2}{(J-1)^2 np_j} \right] \right.$$
$$\left. - \frac{np_k^2}{(np_k + (J-1)\lambda)^3} \frac{\sigma_k^2}{np_k} \right) \stackrel{!}{=} 0$$

The sum notation of the FOC reveals that the smoothing parameter $\lambda$ is non-trivially intertwined across the reference groups. This implies that a closed-form solution exists only in special cases, e.g. for a balanced design when $p_k = 1/J$ for all $k$ or for a design with 2 groups. In general, the FOC is a polynomial equation of any order between 1 and $3(J-1)+1$. This means that already for some designs with more than two groups, one has to solve a polynomial equation of order larger than 4. According to the Abel-Ruffini theorem, there is no guarantee that a solution in radicals exists for polynomial equations of order five and higher with arbitrary coefficients. In these cases, one has to solve the FOC numerically and find the global minimum.

### A.5.3. Generalized Ridge Regression

For the generalized ridge regression (GRR), the restrictions imposed on the smoothing parameters are $\lambda_{kj} = \lambda_k$. This yields an estimator:

$$\hat{\mu}_k^{GRR}(\lambda_k) = \frac{n_k}{n_k + (J-1)\lambda_k} \bar{Y}_k + \frac{(J-1)\lambda_k}{n_k + (J-1)\lambda_k} \sum_{j \neq k} \frac{\hat{\mu}_j}{J-1},$$

which can be rewritten in the following weighted form:

$$\hat{\mu}_k^{GRR}(\omega_k) = (1-\omega_k)\bar{Y}_k + \omega_k \sum_{j \neq k} \frac{\hat{\mu}_j}{J-1}.$$

The GRR estimator depends on the weights within its own reference category $k$. Therefore, optimization of the parameter vector MSE can be done group by group and is invariant to any MSE weighting. Under the choice of $\hat{\mu}_j = \bar{Y}_j$, the MSE of the leading term of a first-order approximation of the GRR estimator takes form[17]:

$$MSE(\hat{\mu}_k^{GRR}(\omega_k)) \approx \omega_k^2 \left[\sum_{j \neq k} \frac{\mu_j}{(J-1)} - \mu_k\right]^2 + (1-\omega_k)^2 \frac{\sigma_k^2}{np_k} + \omega_k^2 \sum_{j \neq k} \frac{\sigma_j^2}{(J-1)^2 np_j}.$$

Optimal solution for $\omega_k$ is:

$$\omega_k^* = \frac{\frac{\sigma_k^2}{np_k}}{\frac{\sigma_k^2}{np_k} + \sum_{j \neq k} \frac{\sigma_j^2}{(J-1)^2 np_j} + \left[\sum_{j \neq k} \frac{\mu_j}{J-1} - \mu_k\right]^2}.$$

### A.6. Proof of Lemma 4.1

**Proof:** $\omega_{kj}^f = O_p(n^{-1})$ if $\{F_n\} \in S(\boldsymbol{\delta}, V_0) \cup S(\infty, V_0)$:

$$n\omega_{kj}^f = \frac{\lambda_{kj}}{n_k/n + \sum_{l \neq k} \lambda_{kl}/n} \xrightarrow{p} \frac{\lambda_{kj}}{p_k} = O(1)$$

by WULLN for $n_k/n$, continuous mapping and assuming $\lambda_{kj}$ fixed. $w_{kk}^f$ follows by definition.

**Proof:** $\omega_{kj}^* \to \bar{w}_{kj} = \frac{\gamma_j(1+\boldsymbol{\delta}'\Delta_k' diag(\boldsymbol{\gamma})\Delta_j\boldsymbol{\delta})}{\boldsymbol{\gamma}'\boldsymbol{\iota}_J + \frac{1}{2}\boldsymbol{\delta}'\Delta' M_1 \Delta\boldsymbol{\delta}}$ if $\{F_n\} \in S(\boldsymbol{\delta}, V_0)$. Use the closed-form in (3.5) and continuity together with $\sqrt{n}\Delta_k\boldsymbol{\mu} \to \Delta_k\boldsymbol{\delta}$ and $\sqrt{n}\Delta\boldsymbol{\mu} \to \Delta\boldsymbol{\delta}$.

**Proof:** $\omega_{kj}^* \to \bar{w}_{kj} = \gamma_j \frac{\boldsymbol{\mu}'\Delta_k' diag(\boldsymbol{\gamma})\Delta_j\boldsymbol{\mu}}{\frac{1}{2}\boldsymbol{\mu}'\Delta' M_1 \Delta\boldsymbol{\mu}}$ if $\{F_n\} \in S(\infty, V_0)$. Follows from dividing by $n$ and taking simple limits, i.e.

$$\omega_{kj}^* = \frac{\gamma_j(1/n + \boldsymbol{\mu}'\Delta_k' diag(\boldsymbol{\gamma})\Delta_j\boldsymbol{\mu})}{\boldsymbol{\gamma}'\boldsymbol{\iota}_J/n + \frac{1}{2}\boldsymbol{\mu}'\Delta' M_1 \Delta\boldsymbol{\mu}} \to 2\gamma_j \frac{\boldsymbol{\mu}'\Delta_k' diag(\boldsymbol{\gamma})\Delta_j\boldsymbol{\mu}}{\boldsymbol{\mu}'\Delta' M_1 \Delta\boldsymbol{\mu}}$$

which exists as $\{F_n\} \in S(\infty, V_0)$.

**Proof:** $\hat{\omega}_{kj} \xrightarrow{d} w_{kj}^a = \frac{\gamma_j(1+(\mathbf{Z}+\boldsymbol{\delta})'\Delta_k diag(\boldsymbol{\gamma})\Delta_j(\mathbf{Z}+\boldsymbol{\delta}))}{\boldsymbol{\gamma}'\boldsymbol{\iota}_J + \frac{1}{2}(\mathbf{Z}+\boldsymbol{\delta})'\Delta' M_1 \Delta(\mathbf{Z}+\boldsymbol{\delta})}$ if $\{F_n\} \in S(\boldsymbol{\delta}, V_0)$. Take $\hat{\omega}_{kj}$ according to (3.6). Note that $\hat{\boldsymbol{\gamma}} \xrightarrow{p} \boldsymbol{\gamma}$ and thus $\hat{M}_1 \xrightarrow{p} M_1$. Additionally $\sqrt{n}(\hat{\mu}_k - \hat{\mu}_j) =$ $\sqrt{n}(\hat{\mu}_k - \mu_k) - \sqrt{n}(\hat{\mu}_j - \mu_j) + \sqrt{n}(\mu_k - \mu_j) \xrightarrow{d} Z_k - Z_j + \delta_k - \delta_j$ since $\{F_n\} \in S(\boldsymbol{\delta}, V_0)$. Similarly $\sqrt{n}\Delta_k\hat{\boldsymbol{\mu}} \xrightarrow{d} \Delta_k(\mathbf{Z} + \boldsymbol{\delta})$ and $\sqrt{n}\Delta\hat{\boldsymbol{\mu}} \xrightarrow{d} \Delta(\mathbf{Z} + \boldsymbol{\delta})$. The rest follows from continuity of $\hat{\omega}_{kj}$.

---

[17]For more details about the first-order approximation please refer to Section 3.

**Proof:** $\hat{\omega}_{kj} \xrightarrow{p} \bar{w}_{kj}$ if $\{F_n\} \in S(\infty, V_0)$. Note that $\hat{\boldsymbol{\mu}} \xrightarrow{p} \boldsymbol{\mu}$, $\hat{\boldsymbol{\gamma}} \xrightarrow{p} \boldsymbol{\gamma}$ and thus $\hat{M}_1 \xrightarrow{p} M_1$. Thus by continuous mapping

$$\hat{\omega}_{kj} = \frac{\hat{\gamma}_j(1/n + \hat{\boldsymbol{\mu}}'\Delta_k' diag(\hat{\boldsymbol{\gamma}})\Delta_j\hat{\boldsymbol{\mu}})}{\hat{\boldsymbol{\gamma}}'\boldsymbol{\iota}_J/n + \frac{1}{2}\hat{\boldsymbol{\mu}}'\Delta'\hat{M}_1\Delta\hat{\boldsymbol{\mu}}} \xrightarrow{p} 2\gamma_j \frac{\boldsymbol{\mu}'\Delta_k' diag(\boldsymbol{\gamma})\Delta_j\boldsymbol{\mu}}{\boldsymbol{\mu}'\Delta' M_1 \Delta\boldsymbol{\mu}}$$

which exists as $\{F_n\} \in S(\infty, V_0)$.

## A.7. Proof of Theorem 4.1

**Proof:** $\sqrt{n}(\hat{\mu}_k^{PCS}(\boldsymbol{\omega}_k^f) - \mu_k - B_{1k}(\boldsymbol{\omega}_k^f)) \xrightarrow{d} Z_k \sim \mathcal{N}\left(0, \frac{\sigma_k^2}{p_k}\right)$ if $\{F_n\} \in S(\boldsymbol{\delta}, V_0) \cup S(\infty, V_0)$. By definition of the PCS and using fixed weights we have

$$\sqrt{n}(\hat{\mu}_k^{PCS}(\boldsymbol{\omega}_k^f) - \mu_k) = \sqrt{n}\sum_{j=1}^J \omega_{kj}^f(\hat{\mu}_j - \mu_j) + \sqrt{n}\sum_{j=1}^J \omega_{kj}^f(\mu_j - \mu_k)$$

Using Lemma 4.1 together with $\sqrt{n}(\hat{\mu}_j - \mu_j) = O_p(1)$ for all $\{F_n\}$ we have that

$$\sqrt{n}(\hat{\mu}_k^{PCS}(\boldsymbol{\omega}_k^f) - \mu_k - \sum_{j=1}^J \omega_{kj}^f(\mu_j - \mu_k)) = \sqrt{n}(\hat{\mu}_k - \mu_k) + o_p(1)$$
$$\xrightarrow{d} Z_k$$

**Proof:** $\sqrt{n}(\hat{\mu}_k^{PCS}(\boldsymbol{\omega}_k^*) - \mu_k - B_{2k}(\boldsymbol{\omega}_k^*)) \xrightarrow{d} \mathcal{N}\left(0, \sum_{j=1}^J \bar{\omega}_{kj}^2 \frac{\sigma_j^2}{p_j}\right)$ if $\{F_n\} \in S(\boldsymbol{\delta}, V_0) \cup S(\infty, V_0)$. Using the definition from the PCS, the CLT for $\sqrt{n}(\hat{\mu}_j - \mu_j)$ together

with Lemma 4.1 yields

$$\sqrt{n}(\hat{\mu}_k^{PCS}(\boldsymbol{\omega}_k^*) - \mu_k - \sum_{j=1}^{J} \omega_{kj}^*(\mu_j - \mu_k)) = \sqrt{n} \sum_{j=1}^{J} \omega_{kj}^*(\hat{\mu}_j - \mu_j)$$

$$= \sqrt{n} \sum_{j=1}^{J} \bar{\omega}_{kj}(\hat{\mu}_j - \mu_j) + o(1)$$

$$\xrightarrow{d} \sum_{j=1}^{J} \bar{\omega}_{kj} Z_j$$

with the final quantity being distributed $\mathcal{N}(0, \sum_{j=1}^{J} \bar{\omega}_{kj}^2 \sigma_j^2 / p_j)$ since $Z_j, Z_k$ are asymptotically independent for all $j \neq k$ due to the orthogonality of the groups.

**Proof:** $\sqrt{n}(\hat{\mu}_k^{PCS}(\hat{\boldsymbol{\omega}}_k) - \mu_k) \xrightarrow{d} \sum_{j=1}^{J} \omega_{kj}^a Z_j + \sum_{j=1}^{J} \omega_{kj}^a (\delta_j - \delta_k)$ if $\{F_n\} \in S(\boldsymbol{\delta}, V_0)$. Rewriting the PCS in the usual manner yields

$$\sqrt{n}(\hat{\mu}_k^{PCS}(\hat{\boldsymbol{\omega}}_k) - \mu_k) = \sqrt{n} \sum_{j=1}^{J} \hat{\omega}_{kj}(\hat{\mu}_j - \mu_j) + \sqrt{n} \sum_{j=1}^{J} \hat{\omega}_{kj}(\mu_j - \mu_k)$$

$$\xrightarrow{d} \sum_{j=1}^{J} \omega_{kj}^a Z_j + \sum_{j=1}^{J} \omega_{kj}^a (\delta_j - \delta_k)$$

where convergence in distribution follows from joint convergence of the $\hat{\omega}_{kj}$'s and $\sqrt{n}(\hat{\mu}_j - \mu_j)$'s as they are continuous functions of the same random normal vector and using the distributional Lemma for the weights for $\{F_n\} \in S(\boldsymbol{\delta}, V_0)$ and $\sqrt{n}(\mu_j - \mu_k) \to \delta_k - \delta_j$ by definition of sequences in $S(\boldsymbol{\delta}, V_0)$.

**Proof:** $\sqrt{n}(\hat{\mu}_k^{PCS}(\hat{\boldsymbol{\omega}}_k) - \mu_k - B_{3k}(\bar{\boldsymbol{\omega}}_k)) \xrightarrow{d} Z_k \sim \mathcal{N}\left(0, \frac{\sigma_k^2}{p_k}\right)$ if $\{F_n\} \in S(\infty, V_0)$. By Lemma 4.1, $\hat{\omega}_{kj} \xrightarrow{p} \bar{\omega}_{kj}$ as $\{F_n\} \in S(\infty, V_0)$. Rewriting the PCS yields

$$\hat{\mu}_k^{PCS}(\hat{\boldsymbol{\omega}}_k) - \mu_k = \sum_{j=1}^{J} \hat{\omega}_{kj}(\hat{\mu}_j - \mu_j + \mu_j - \mu_k)$$

$$= \sum_{j=1}^{J} \hat{\omega}_{kj}(\hat{\mu}_j - \mu_j) + \sum_{j=1}^{J} \bar{\omega}_{kj}(\mu_j - \mu_k) + \sum_{j=1}^{J} (\hat{\omega}_{kj} - \bar{\omega}_{kj})(\mu_j - \mu_k)$$

or equivalently

$$\sqrt{n}(\hat{\mu}_k^{PCS}(\hat{\boldsymbol{\omega}}_k) - \mu_k - \sum_{j \neq k} \bar{\omega}_{kj}(\mu_j - \mu_k)) = \sum_{j=1}^{J}(\hat{\omega}_{kj} - \bar{\omega}_{kj})\sqrt{n}(\hat{\mu}_j - \mu_j)$$
$$+ \sum_{j=1}^{J} \bar{\omega}_{kj}\sqrt{n}(\hat{\mu}_j - \mu_j) + \sum_{j \neq k} \sqrt{n}(\hat{\omega}_{kj} - \bar{\omega}_{kj})(\mu_j - \mu_k)$$
$$= \sum_{j=1}^{J} \bar{\omega}_{kj}\sqrt{n}(\hat{\mu}_j - \mu_j) + \sum_{j \neq k} \sqrt{n}(\hat{\omega}_{kj} - \bar{\omega}_{kj})(\mu_j - \mu_k) + o_p(1).$$

The right hand side is asymptotically normal as the components are stabilizing transformations of continuous functions of the same random normal vector. In terms of its asymptotic variance, one can either show the equivalence to $Z_k$ using the delta method or more simple by Theorem 5.1. It implies that as $||\Delta\boldsymbol{\delta}||_\infty \to \infty$, the PCS risk is converging to the OLS. Since both estimators are asymptotically normal, the asymptotic variances have to coincide.

## A.8. Proof of Theorem 5.1 and Corollary 5.1

**Proof:** Let $\{F_n\} \in S(\boldsymbol{\delta}, V_0)$. The plugin weights are given by

$$\hat{\omega}_{kj} = \frac{\hat{\gamma}_j + n\sum_{m=1}^{J}(\hat{\mu}_k - \hat{\mu}_m)(\hat{\mu}_j - \hat{\mu}_m)\hat{\gamma}_j\hat{\gamma}_m}{\sum_{l=1}^{J}\hat{\gamma}_l + 0.5n\sum_{l=1}^{J}\sum_{m=1}^{J}(\hat{\mu}_l - \hat{\mu}_m)^2\hat{\gamma}_l\hat{\gamma}_m}.$$

which by Lemma 4.1 converge in distribution, i.e.

$$\hat{\omega}_{kj} \xrightarrow{d} w_{kj}^a = \frac{\gamma_j + \sum_{m=1}^{J}(Z_k - Z_m + \delta_k - \delta_m)(Z_j - Z_m + \delta_j - \delta_m)\gamma_j\gamma_m}{d_0}$$

with $d_0 = \boldsymbol{\gamma}'\boldsymbol{\iota}_J + \frac{1}{2}(\mathbf{Z} + \boldsymbol{\delta})'\Delta'M_1\Delta(\mathbf{Z} + \boldsymbol{\delta})$. By Theorem 4.1, the distributional limit for the PCS under $\{F_n\} \in S(\boldsymbol{\delta}, V_0)$ is given by

$$\sqrt{n}(\hat{\mu}_k^{PCS}(\hat{\boldsymbol{\omega}}_k) - \mu_k) \xrightarrow{d} \sum_{j=1}^{J}\omega_{kj}^a Z_j + \sum_{j=1}^{J}\omega_{kj}^a(\delta_j - \delta_k) = \sum_{j=1}^{J}\omega_{kj}^a(Z_j - Z_k + \delta_j - \delta_k) + Z_k \equiv \psi_k$$

since $\sum_{j=1}^{J} \omega_{kj}^a = 1$ for all $k$. By Lemma 1 of Hansen (2016a), the asymptotic weighted MSE criterion then yields

$$\rho(\hat{\boldsymbol{\mu}}^{PCS}(\hat{\boldsymbol{\omega}}), \boldsymbol{\mu}) = \sum_{k=1}^{J} \gamma_k E[\psi_k^2]$$

$$= E[\sum_{k=1}^{J} \sum_{j=1}^{J} \sum_{l=1}^{J} \gamma_k \gamma_j \gamma_l (Z_k - Z_j + \delta_k - \delta_j)(Z_k - Z_l + \delta_k - \delta_l)/d_0^2]$$

$$- 2E[\sum_{k=1}^{J} \sum_{j=1}^{J} \gamma_k \gamma_j (Z_k - Z_j + \delta_k - \delta_j) Z_k / d_0] + E[\sum_{k=1}^{J} \gamma_k Z_k^2]$$

$$\equiv E[A] - 2E[B] + \rho(\hat{\boldsymbol{\mu}}, \boldsymbol{\mu})$$

with

$$A = (\mathbf{Z} + \boldsymbol{\delta})' \Delta' M_2 \Delta (\mathbf{Z} + \boldsymbol{\delta}) / d_0^2$$

$$M_2 = diag(\boldsymbol{\gamma}) \otimes \boldsymbol{\gamma}\boldsymbol{\gamma}'$$

$$B = (\mathbf{Z} + \boldsymbol{\delta})' \Delta' M_3 \mathbf{Z} / d_0$$

$$M_3 = diag(\boldsymbol{\gamma}) \otimes \boldsymbol{\gamma}$$

To further simplify $E[B]$ we use a multivariate version of Stein's Lemma given by Lemma 2 in Hansen (2016a) which yields

$$E[B] = E[\eta(\mathbf{Z} + \boldsymbol{\delta})' \Delta' M_3 \mathbf{Z}] = E\left[tr\left(\frac{\partial}{\partial \mathbf{x}} \eta(\mathbf{Z} + \boldsymbol{\delta})' \Delta' M_3 V_0\right)\right]$$

with $\eta(\mathbf{x}) = \mathbf{x}/(\boldsymbol{\gamma}' \boldsymbol{\iota}_J + 0.5 \mathbf{x}' \Delta' M_1 \Delta \mathbf{x})$ and derivative

$$\frac{\partial}{\partial \mathbf{x}} \eta(\mathbf{x})' = \frac{1}{d_0} I_J - \frac{\Delta' M_1 \Delta}{d_0^2} \mathbf{x}\mathbf{x}'$$

and hence

$$E[B] = tr(\Delta' M_3 V_0) E\left[\frac{1}{d_0}\right] - E\left[\frac{tr(\Delta' M_1 \Delta (\mathbf{Z} + \boldsymbol{\delta})(\mathbf{Z} + \boldsymbol{\delta})' \Delta' M_3 V_0)}{d_0^2}\right]$$

$$= \boldsymbol{\gamma}' \boldsymbol{\iota}_J tr(\Delta' M_3 V_0) E\left[\frac{1}{d_0^2}\right] + \frac{1}{2} tr(\Delta' M_3 V_0) E\left[\frac{(\mathbf{Z} + \boldsymbol{\delta})' \Delta' M_1 \Delta (\mathbf{Z} + \boldsymbol{\delta})}{d_0^2}\right]$$

$$- E\left[\frac{(\mathbf{Z} + \boldsymbol{\delta})' \Delta' M_3 V_0 \Delta' M_1 \Delta (\mathbf{Z} + \boldsymbol{\delta})}{d_0^2}\right].$$

Since $\boldsymbol{\gamma}'\boldsymbol{\iota}_J = tr(V_0^{-1})$, plugging in and bringing the terms together with $E[A]$ yields the following asymptotic risk:

$$\rho(\hat{\boldsymbol{\mu}}^{PCS}(\hat{\boldsymbol{\omega}}), \boldsymbol{\mu}) = \rho(\hat{\boldsymbol{\mu}}, \boldsymbol{\mu}) + E\left[\frac{(\mathbf{Z}+\boldsymbol{\delta})'\Delta'C\Delta(\mathbf{Z}+\boldsymbol{\delta})}{(tr(V_0^{-1}) + \frac{1}{2}(\mathbf{Z}+\boldsymbol{\delta})'\Delta'M_1\Delta(\mathbf{Z}+\boldsymbol{\delta}))^2}\right]$$
$$- 2tr(V_0^{-1})tr(\Delta'M_3V_0)E\left[\frac{1}{(tr(V_0^{-1}) + \frac{1}{2}(\mathbf{Z}+\boldsymbol{\delta})'\Delta'M_1\Delta(\mathbf{Z}+\boldsymbol{\delta}))^2}\right]$$

with $C = M_2 - tr(\Delta'M_3V_0)M_1 + 2M_3V_0\Delta'M_1$. The corollary then follows from $C$ being negative semidefinite if $J > 3$. We show that for $J > 3$, $-C$ is positive semidefinite. Some algebra yields the following characterization for $-C$:

$$-C_{ij} = \begin{cases} \gamma_i(2J-7)\sum_{l \neq i}\gamma_l \sum_{m=1}^{J}\gamma_m & \text{if } i = j \\ -\gamma_i\gamma_j(2J-7)\sum_{l=1}^{J}\gamma_l & \text{if } i \neq j. \end{cases}$$

Due to positivity of the $\gamma_j$'s, the diagonal elements of $-C$ are strictly positive if $(2J-7) > 3$. Thus, a sufficient condition for $-C$ being positive semidefinite is (absolute) diagonal dominance, i.e. $|-C_{ii}| \geq \sum_{j \neq i}|-C_{ij}|$ which yields

$$\gamma_i(2J-7)\sum_{l \neq i}\gamma_l \sum_{m=1}^{J}\gamma_m \geq \sum_{j \neq i}\gamma_i\gamma_j(2J-7)\sum_{l=1}^{J}\gamma_j \quad \Leftrightarrow 0 \geq 0$$

which proofs the sufficiency.

## A.9. Supplementary Material for Section 7.2

Table A.1: Summary Statistics of the Card and Krueger (1994) Data

| Chain | NJ (treated) | | | | | | PEN (control) | | | | | |
| --- | --- | --- | --- | --- | --- | --- | --- | --- | --- | --- | --- | --- |
| | Before | | | After | | | Before | | | After | | |
| | $\hat{\mu}$ | $\sigma^2$ | $n$ | $\hat{\mu}$ | $\sigma^2$ | $n$ | $\hat{\mu}$ | $\sigma^2$ | $n$ | $\hat{\mu}$ | $\sigma^2$ | $n$ |
| All chains | 20.44 | 82.92 | 321 | 21.03 | 86.36 | 319 | 23.33 | 140.57 | 77 | 21.17 | 68.5 | 77 |
| Burger King | 22.16 | 61.95 | 131 | 23.63 | 70.63 | 131 | 29.42 | 182.81 | 33 | 26.22 | 50.31 | 35 |
| KFC | 12.79 | 21.83 | 67 | 13.73 | 39.60 | 68 | 10.71 | 7.83 | 12 | 13.00 | 11.59 | 12 |
| Roys | 23.14 | 109.36 | 81 | 21.73 | 89.30 | 78 | 19.74 | 32.96 | 17 | 15.81 | 43.89 | 17 |
| Wendys | 22.08 | 79.99 | 42 | 23.40 | 96.64 | 42 | 24.12 | 61.20 | 15 | 22.10 | 39.35 | 13 |